\documentclass[3p,number,review,sort&compress]{elsarticle}

\usepackage{psfrag}
\usepackage{amsmath,amssymb}
\usepackage{dsfont}
\usepackage{color}
\usepackage{nicefrac}
\usepackage{appendix}
\usepackage{cprotect}
\usepackage{multirow}

\begin{document}


\title{Chaos in Coupled Heteroclinic Cycles and its Piecewise-Constant Representation}

\author[up,hse]{Arkady Pikovsky}%
\ead{pikovsky@uni-potsdam.de}
\author[tech]{Alexander Nepomnyashchy}
 \ead{nepom@math.technion.ac.il}
\address[up]{%
 Institute for Physics and Astronomy,  University of Potsdam, 
 Karl-Liebknecht Str. 24/25,
 14476 Potsdam, Germany 
}%
\address[hse]{National Research University Higher School of Economics,  Nizhny Novgorod, Russia}
\address[tech]{ Department of Mathematics, Technion - Israel Institute of Technology, Haifa, Israel}
\begin{abstract}
We consider two stable heteroclinic cycles rotating in opposite directions, coupled via diffusive terms. A complete synchronization in this system is impossible, and numerical exploration shows that chaos is abundant
at low levels
of coupling. With increase of coupling strength, several symmetry-changing transitions are observed, and finally a stable periodic orbit appears via an inverse period-doubling cascade. To reveal the behavior at extremely small couplings, a piecewise-constant model for the dynamics is suggested. Within this model we construct a Poincar\'e map for a chaotic state
numerically, it appears to be an expanding non-invertable circle map thus confirming abundance of chaos in the small coupling limit. We also show that
within the piecewise-constant description, there is a set of periodic solutions with different phase shifts between subsystems, due to dead zones in the coupling.
\end{abstract}

\maketitle

\section{Introduction}
\label{sec:intro}

A heteroclinic cycle is an interesting object in nonlinear dynamics, attracted large of attention recently. It was probably first described by May and Leonard \cite{May-Leonard-75}
in the context of a generalization of the Lotka-Volterra model of species competition. One
of the early examples is also the dynamics of three unstable modes in a convection
problem in a rotating fluid layer heated from below~\cite{MR1708859}. From
the mathematical viewpoint, a stable heteroclinic cycle is a robust object, demonstrating
oscillations the period of which grows in time, and tends to infinity~\cite{MR913462,MR1437986}.
Among recent applications we mention the concept of winnerless competition in neurosciences, based on a representation of certain dynamical patterns as vicinities of a heteroclinic cycle~\cite{MR2063888,MR2513781}; studies of the dynamics of spin-torque nano-spintronic oscillators~\cite{PhysRevB.84.104414}; synchronization dynamics
of coupled oscillators~\cite{MR4030395}. In the simplest setup, a heteroclinic cycle includes three interacting ``species'', but recently generalizations to larger networks have been also investigated~\cite{TACHIKAWA2007374,10.1143/PTP.109.133,voit2019dynamics,Voit_etal-20}.

In this paper, we consider a simple system of two coupled
subsystems, each of which possesses a stable heteroclinic cycle.
For brevity, we will speak on ``coupled heteroclinic cycles''. A variant with a symmetric (or nearly symmetric) coupling has been studied in \cite{PhysRevE.85.016215}. There, several synchronous and quasiperiodic regimes have been reported. Here we consider a strongly antisymmetric coupling of two cycles. We assume, that we have two cycles oscillating
\textit{in opposite directions}. Namely, while in one cycle we have a sequence of states $1\to 2\to 3 \to 1\to 2\ldots$, in another cycle we have a sequence of states $1\to 3 \to 2 \to 1 \to 3 \ldots$. The coupling is an attraction of the corresponding states in two cycles. Because of opposite rotations, a completely synchronous state, where the states of the coupled variables  in two subsystems coincide (and thus the coupling diffusion terms vanish),  is impossible (even if all parameters of the cycles are the same, as we assume below). We will however show, that partially synchronized periodic regimes, where
non-interacting variables coincide, can appear and are stable for large coupling. Our main finding is that chaos dominates the dynamics at small coupling. We first show this numerically for the full system of equations. Then, we develop an approximate approach where the dynamics is approximated by piecewise-constant functions. This approach captures the properties of coupled heteroclinic cycles in the limit of small coupling.

The paper is organized as follows. We discuss the basic heteroclinic cycle in Section~\ref{sec:hc}. Coupled heteroclinic cycles are introduced in Section~\ref{sec:chc}. The coupled system possesses several symmetries, which we discuss in Section~\ref{sec:ism}. Numerical exploration of coupled cycles is performed in Section~\ref{sec:nexp}. In Section \ref{sec:pla} we introduce 
the piecewise-constant model of the cycle dynamics, which is then extended to coupled units
in Section~\ref{sec:pc}. Chaotic and periodic regimes within the piecewise-constant model are explored in Section~\ref{sec:plach}, and compared with the regimes in the full system in Section~\ref{sec:ecl}. We conclude
with discussion of the results in Section~\ref{sec:concl}.

\section{Heteroclinic cycle}
\label{sec:hc}

We discuss here briefly the canonical model of a heteroclinic cycle,
introduced by May and Leonard~\cite{May-Leonard-75}:
\begin{equation}
\begin{aligned}
\dot u&=u(1-u-\alpha v-\beta w)\;,\\
\dot v&=v(1-v-\alpha w-\beta u)\;,\\
\dot w&=w(1-w-\alpha u-\beta v)\;.
\end{aligned}
\label{eq:hc1}
\end{equation}
We assume below that $\alpha>1$ and $0<\beta<1$. The crucial parameter 
here is the sum $\alpha+\beta$. For $\alpha+\beta>2$, a stable heteroclinic cycle exists, in which a trajectory consequently approaches steady states $(1,0,0)\to(0,1,0)\to(0,0,1)\to(1,0,0)\to\ldots$ as time grows. The time intervals 
that a trajectory spends in a vicinity of these states grow exponentially.

\begin{figure}[!htb]
\centering
\includegraphics[width=0.6\textwidth]{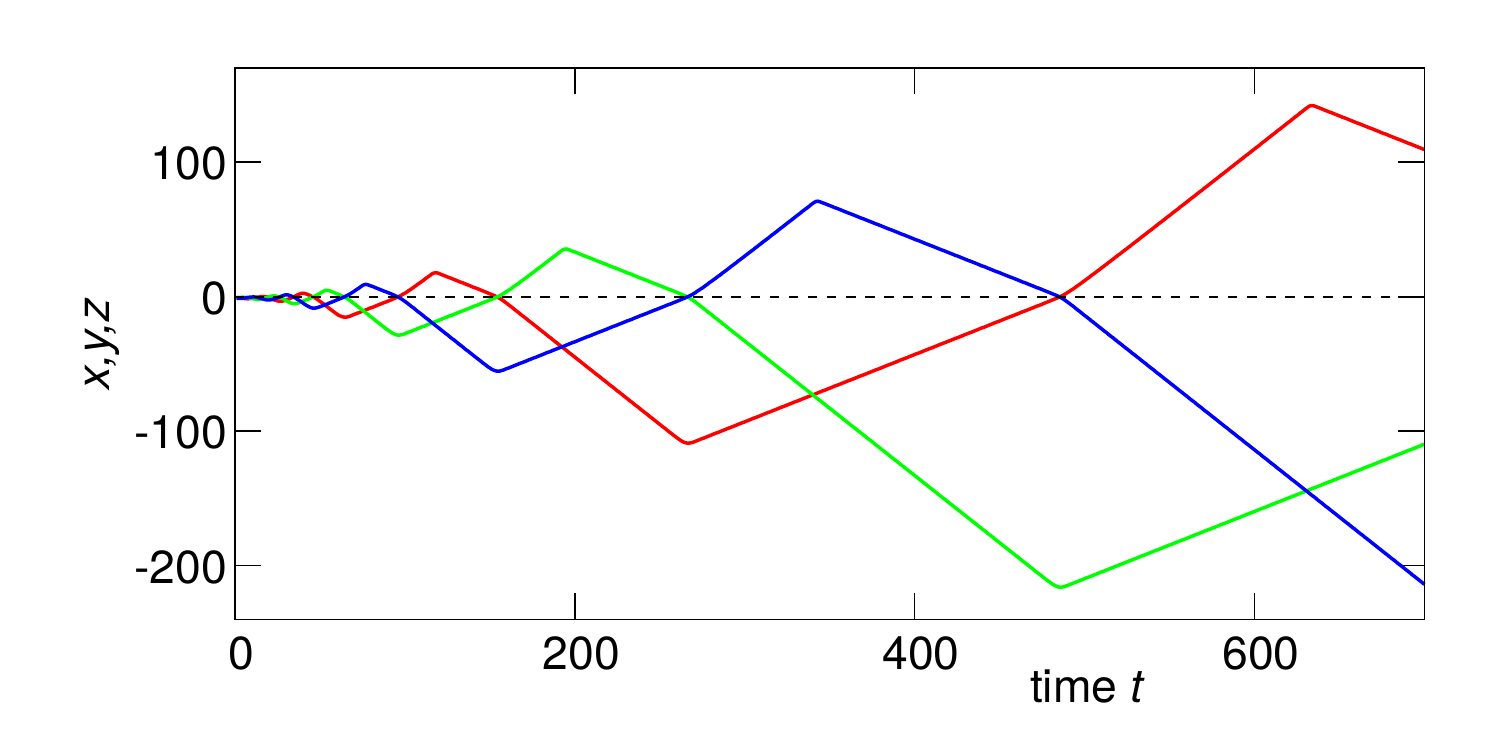}
\caption{A trajectory in systems \eqref{eq:hc1},\eqref{eq:tr}. 
Colors: Red: $x(t)$, green: $y(t)$, blue: $z(t)$. Large positive values
of $x$ correspond to $u\approx 1$, and large negative values of $x$ correspond to $u\approx 0$;
similarly for other variables.}
\label{fig:c}
\end{figure}

For an illustration and for further analysis below, it is convenient
to introduce new variables allowing for a better resolution of vicinities
of the steady states, stable and unstable manifolds of which constitute the heteroclinic cycle. These variables are also suitable for numerical
integration and for a piecewise-constant approximation (see Section~\ref{sec:pla} below).
These new variables read
\begin{equation}
    x=\ln\frac{u}{1-u}\;,\quad y=\ln\frac{v}{1-v}\;,\quad z=\ln\frac{w}{1-w}\;.
\label{eq:xyzvar}    
\end{equation}
The equations \eqref{eq:hc1} in these variables take the form
\begin{equation}
\begin{aligned}
\dot x &=1-\alpha \frac{1+e^x}{1+e^{-y}}-\beta \frac{1+e^x}{1+e^{-z}}\;,\\
\dot y&=1-\alpha \frac{1+e^y}{1+e^{-z}}-\beta \frac{1+e^y}{1+e^{-x}}\;,\\
\dot z&=1-\alpha \frac{1+e^z}{1+e^{-x}}-\beta \frac{1+e^z}{1+e^{-y}}\;.
\end{aligned}
\label{eq:tr}
\end{equation}
We illustrate a trajectory approaching the heteroclinic cycle
in Fig.~\ref{fig:c}.

\section{Coupled heteroclinic cycles}
\label{sec:chc}

Here we introduce our basic model of two interacting heteroclinic cycles.
In contradistinction to the previously explored case of a nearly symmetric
coupling~\cite{PhysRevE.85.016215}, we consider an asymmetric coupling: we assume that two heteroclinic cycles of type \eqref{eq:hc1}
\textit{rotate in opposite directions}. Namely, while in cycle 1 with variables $u_1,v_1,w_1$ the excitations are according to the rule
$u_1\to v_1\to w_1\to u_1\ldots$, in cycle 2 the excitations
go as $u_2\to w_2\to v_2\to u_2\to\ldots$. The coupled variables
are pairs $(u_1,u_2)$, $(v_1,v_2)$, and $(w_1,w_2)$. The equations of our model read
\begin{equation}
\begin{aligned}
\dot u_1&=u_1(1-u_1-\alpha v_1-\beta w_1)+D(u_2-u_1)\;,\\
\dot v_1&=v_1(1-v_1-\alpha w_1-\beta u_1)+D(v_2-v_1)\;,\\
\dot w_1&=w_1(1-w_1-\alpha u_1-\beta v_1)+D(w_2-w_1)\;,\\
\dot u_2&=u_2(1-u_2-\beta v_2-\alpha w_2)+D(u_1-u_2)\;,\\
\dot v_2&=v_1(1-v_2-\beta w_2-\alpha u_2)+D(v_1-v_2)\;,\\
\dot w_2&=w_1(1-w_2-\beta u_2-\alpha v_2)+D(w_1-w_2)\;.
\end{aligned}
\label{eq:2c}
\end{equation}
One can see that because of exchange $\alpha\leftrightarrow\beta$, the cycles
in subsystems 1 and 2 rotate in opposite directions. Parameter $D$ describes coupling between subsystems. We expect that already a small coupling will significantly influence the dynamics, so $D$ will be mostly small. Therefore,
it is convenient to represent this parameter as $D=\exp[-Q]$, so that large values of $Q$ correspond to a weak coupling. Then, adopting
the variables \eqref{eq:xyzvar}, we can rewrite system \eqref{eq:2c} as
\begin{equation}
\begin{aligned}
\dot x_1&=1-\alpha \frac{1+e^{x_1}}{1+e^{-{y_1}}}-\beta \frac{1+e^{x_1}}{1+e^{-z_1}}
+e^{-Q}\left[\frac{(1+e^{-x_1})(1+e^{x_1})}{1+e^{-x_2}}-(1+e^{x_1})\right]\;,\\
\dot y_1&=1-\alpha \frac{1+e^{y_1}}{1+e^{-z_1}}-\beta \frac{1+e^{y_1}}{1+e^{-x_1}}
+e^{-Q}\left[\frac{(1+e^{-y_1})(1+e^{y_1})}{1+e^{-y_2}}-(1+e^{y_1})\right]\;,\\
\dot z_1&=1-\alpha \frac{1+e^{z_1}}{1+e^{-x_1}}-\beta \frac{1+e^{z_1}}{1+e^{-{y_1}}}
+e^{-Q}\left[\frac{(1+e^{-z_1})(1+e^{z_1})}{1+e^{-z_2}}-(1+e^{z_1})\right]\;,\\
\dot x_2&=1-\beta \frac{1+e^{x_2}}{1+e^{-y_2}}-\alpha \frac{1+e^{x_2}}{1+e^{-z_2}}
+e^{-Q}\left[\frac{(1+e^{-x_2})(1+e^{x_2})}{1+e^{-x_1}}-(1+e^{x_2})\right]\;,\\
\dot y_2&=1-\beta \frac{1+e^{y_2}}{1+e^{-z_2}}-\alpha \frac{1+e^{y_2}}{1+e^{-x_2}}
+e^{-Q}\left[\frac{(1+e^{-y_2})(1+e^{y_2})}{1+e^{-y_1}}-(1+e^{y_2})\right]\;,\\
\dot z_2&=1-\beta \frac{1+e^{z_2}}{1+e^{-x_2}}-\alpha \frac{1+e^{z_2}}{1+e^{-y_2}}
+e^{-Q}\left[\frac{(1+e^{-z_2})(1+e^{z_2})}{1+e^{-z_1}}-(1+e^{z_2})\right]\;.\\
\end{aligned}
\label{eq:trc}
\end{equation}
While Eqs.~\eqref{eq:2c} are appropriate for an analytical consideration
of regimes where the co-existence state is weakly unstable, Eqs.~\eqref{eq:trc} are suitable for numerical explorations of a deeply
heteroclinic dynamics.

\section{Invariant synchronous manifolds}
\label{sec:ism}
\subsection{Symmetry properties of systems \eqref{eq:2c}, \eqref{eq:trc}}
System \eqref{eq:2c} is invariant with respect to {\em renamings} of variables inside each subsystem:
\begin{equation}
\begin{gathered}
R_1:(u_1,v_1,w_1,u_2,v_2,w_2)\to(v_1,w_1,u_1,v_2,w_2,u_2)\\
\text{and }
R_2:(u_1,v_1,w_1,u_2,v_2,w_2)\to(w_1,u_1,v_1,w_2,u_2,v_2).
\end{gathered}
\label{eq:renam}
\end{equation}
Obviously, $R_2=R_1^{-1}=R_1^2$, $R_1=R_2^{-1}=R_2^2$. Also, system \eqref{eq:2c} is symmetric with respect to three {\em exchange} transformations,
\begin{equation}
\begin{gathered}
T_1:(u_1,v_1,w_1,u_2,v_2,w_2)\to(u_2,w_2,v_2,u_1,w_1,v_1)\;,\\
T_2:(u_1,v_1,w_1,u_2,v_2,w_2)\to(w_2,v_2,u_2,w_1,v_1,u_1)\;,\\
\text{and }
T_3:(u_1,v_1,w_1,u_2,v_2,w_2)\to(v_2,u_2,w_2,v_1,u_1,w_1)\;.
\end{gathered}
\label{eq:exch}
\end{equation}
One can see that $T_1^2=T_2^2=T_3^2=I$, $T_2=R_2T_1$, $T_3=R_1T_1$.

System \eqref{eq:trc} has the same set of symmetries with replacement of $(u_i,v_i,w_i)$ with $(x_i,y_i,z_i)$, $i=1,2$.

The subspaces invariant with respect to transformations $T_1$, $T_2$ and $T_3$ are invariant manifolds for system \eqref{eq:2c}. Let us consider, for instance, i.e., the transformation
$$T_1:(x_1,y_1,z_1,x_2,y_2,z_2)\to(x_2,z_2,y_2,x_1,z_1,y_1).$$
System \eqref{eq:trc} can be written as
$$\dot{x}_1=f(x_1,y_1,z_1,x_2),\;\dot{y}_1=f(y_1,z_1,x_1,y_2),\;\dot{z}_1=f(z_1,x_1,y_1,z_2),$$
$$\dot{x}_2=f(x_2,z_2,y_2,x_1),\;\dot{y}_2=f(y_2,x_2,z_2,y_1),\;\dot{z}_2=f(z_2,y_2,x_2,z_1),$$
where
$$f(x,y,z,u)=1-\alpha\frac{1+e^x}{1+e^{-y}}-\beta\frac{1+e^x}{1+e^{-z}}+e^{-Q}(1+e^x)\frac{e^{-x}-e^{-u}}{1+e^{-u}}\;.$$
Let us define
$$x=\frac{x_1+x_2}{2},\;y=\frac{y_1+z_2}{2},\;z=\frac{z_1+y_2}{2},\;X=\frac{x_2-x_1}{2},\;Y=\frac{z_2-y_1}{2},\;Z=\frac{y_2-z_1}{2}.$$
Then we obtain:
$$\dot{X}=\frac{1}{2}[f(x+X,y+Y,z+Z,x-X)-f(x-X,y-Y,z-Z,x+X)]\;,$$
$$\dot{Y}=\frac{1}{2}[f(y+Y,z+Z,x+X,z-Z)-f(y-Y,z-Z,x-X,z+Z)]\;,$$
$$\dot{Z}=\frac{1}{2}[f(z+Z,x+X,y+Y,y-Y)-f(z-Z,x-X,y-Y,y+Y)]\;,$$
$$\dot{x}=\frac{1}{2}[f(x+X,y+Y,z+Z,x-X)+f(x-X,y-Y,z-Z,x+X)]\;,$$
$$\dot{y}=\frac{1}{2}[f(y+Y,z+Z,x+X,z-Z)+f(y-Y,z-Z,x-X,z+Z)]\;,$$
$$\dot{z}=\frac{1}{2}[f(z+Z,x+X,y+Y,y-Y)+f(z-Z,x-X,y-Y,y+Y)]\;.$$

One can see that the system has a 3-dimensional invariant manifold $X=Y=Z=0$.
The dynamics on that manifold is considered in the next subsection \ref{sec:im}. 
The invariance to transformation $T_1$ corresponds to the reversion symmetry with respect to that manifold:
$$(x,y,z,X,Y,Z)\to(x,y,z,-X,-Y,-Z).$$ 
Therefore, the system can have a symmetric attractor that includes simultaneously both points $(x,y,z,X,Y,Z)$ and $(x,y,z,-X,-Y,-Z)$. Otherwise, it has two asymmetric attractors connected by $T_1$.
\subsection{Invariant manifolds}
\label{sec:im}
System \eqref{eq:2c} has a 3-dimensional invariant manifold
\begin{equation}
M1:\;u_1=u_2\equiv u,\;v_1=w_2\equiv v,\;w_1=v_2\equiv w,
\label{eq:mf1}
\end{equation}
on which the dynamics reduces to a system of 3 equations:
\begin{equation}
\begin{aligned}
\frac{du}{dt}&=u(1-u-\alpha v-\beta w),\\
\frac{dv}{dt}&=v(1-v-\alpha w-\beta u)+D(w-v),\\
\frac{dw}{dt}&=w(1-w-\alpha u-\beta v)+D(v-w).
\end{aligned}
\label{eq:m1}  
\end{equation}
Such a regime can be described as a partially synchronous one. Here,
as one can see from expressions \eqref{eq:mf1}, one pair of interacting variables are identical, while two other
pairs are ``cross-identical''. Due to this, the coupling terms are essential for the dynamics.

Due to the renaming symmetry, system \eqref{eq:2c} has 
 two other invariant manifolds, 
\begin{equation}
M2:\;v_1=v_2\equiv u,\;w_1=u_2\equiv v,\;u_1=w_2\equiv w,\qquad
M3:\;w_1=w_2\equiv u,\;u_1=v_2\equiv v,\;v_1=u_2\equiv w,
\label{eq:mf23}
\end{equation}
where the dynamics is described by exactly the same system 
\eqref{eq:m1}.
\subsection{Stationary points and their stability}
For $D=0$, system \eqref{eq:m1} coincides with \eqref{eq:hc1} and has
five steady states:
\[
(0,0,0),\quad (1,0,0),\quad (a,a,a),\quad(0,1,0),\quad (0,0,1)
\]
where $a=(\alpha+\beta+1)^{-1}$. The first three of these points do
not depend on parameter $D$. For small values of $D$, we 
can approximately find the last two steady states up to the first order in $D$:
\[
(0,1+D\frac{\alpha+\beta-1}{1-\beta},D\frac{-1}{1-\beta}),\quad
(0,D\frac{1}{\alpha-1},1+D\frac{-\alpha+1-\beta}{\alpha-1})
\]
Because $\beta<1$, the first of these steady states lies in a ``forbidden domain''
$w<0$, and only the second of these steady states exists. Thus, a heteroclinic cycle disappears for any small $D$.

For $D=0$, the ``coexistence point'' $(a,a,a)$ loses its stability
at $\alpha+\beta=2$, and the heteroclinic cycle appears. In system
\eqref{eq:2c} the relevant eigenvalues of the coexistence point are
\begin{equation}
\sigma_{2,3}=\frac{\alpha+\beta-2}{2(\alpha+\beta+1)}-D\pm 
i\left[\frac{3(\alpha-\beta)^2}{4(\alpha+\beta+1)^2}
-D^2\right]^{1/2}.
\label{eq:scp}
\end{equation}
Thus, the ``coexistence point" is oscillatory unstable if 
$D<\frac{\alpha+\beta-2}{2(\alpha+\beta+1)}$. For small $D$,
 the shift of the stability threshold is 
$\alpha+\beta-2=6D+O(D^2)$. Furthermore, as is shown in  \ref{sec:hbif}, the transition is now a standard Hopf bifurcation at which
a stable limit cycle appears. In this paper, however,
we will not explore the case where parameters $\alpha$ and $\beta$ are close
to this bifurcation point, rather we focus on a situation where the trajectory
in coupled systems
is close to former
well-developed heteroclinic cycles (i.e. with $\alpha+\beta-2$ not small).

\begin{figure}[!htb]
\centering
\includegraphics[width=0.49\textwidth]{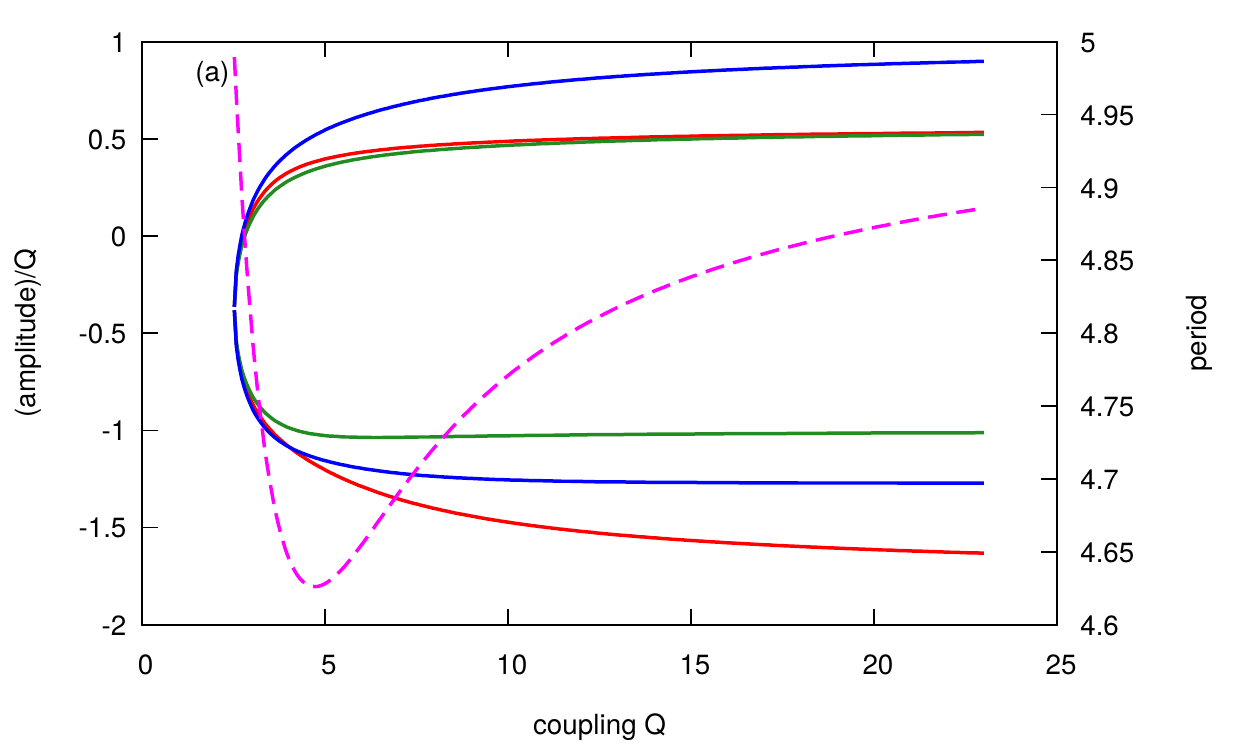}\hfill
\includegraphics[width=0.49\textwidth]{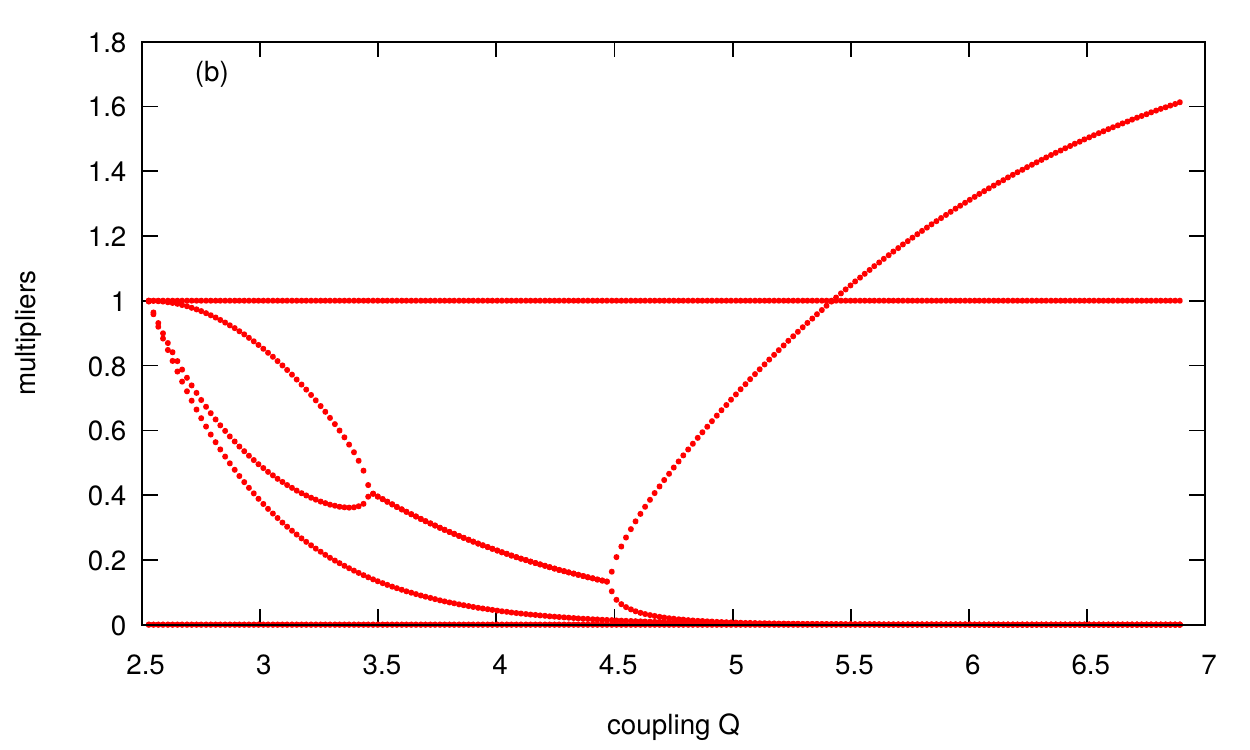}
\caption{Panel (a): period of the cycle (magenta dashed line,
right axis) and maximal and minimal values of the normalized by $Q$ variables $x$ (red), $y$ (green) and $z$ (blue, left axis). 
Panel (b): absolute values
of the multipliers of the cycle. Stability is lost when one real multiplier crosses the border $-1$.}
\label{fig:manM1}
\end{figure}

Below in numerical simulations in Sections~\ref{sec:ism},\ref{sec:nexp} we use 
\begin{equation}
\begin{gathered}
\alpha=1+\Omega\approx 2.324,\quad \beta=1-\Omega^{-1}\approx 0.2451, \\
\text{where}\quad \Omega=1.32471595724475\ldots\quad \text{is the spiral mean }\Omega^3=\Omega+1\;.
\end{gathered}
\label{eq:spm}
\end{equation} 
Furthermore, for better comparison with theory of 
Section~\ref{sec:pc}, we normalize the depicted variables $x,y,z$ by $Q$.

\begin{figure}[!htb]
\centering
\includegraphics[width=0.7\textwidth]{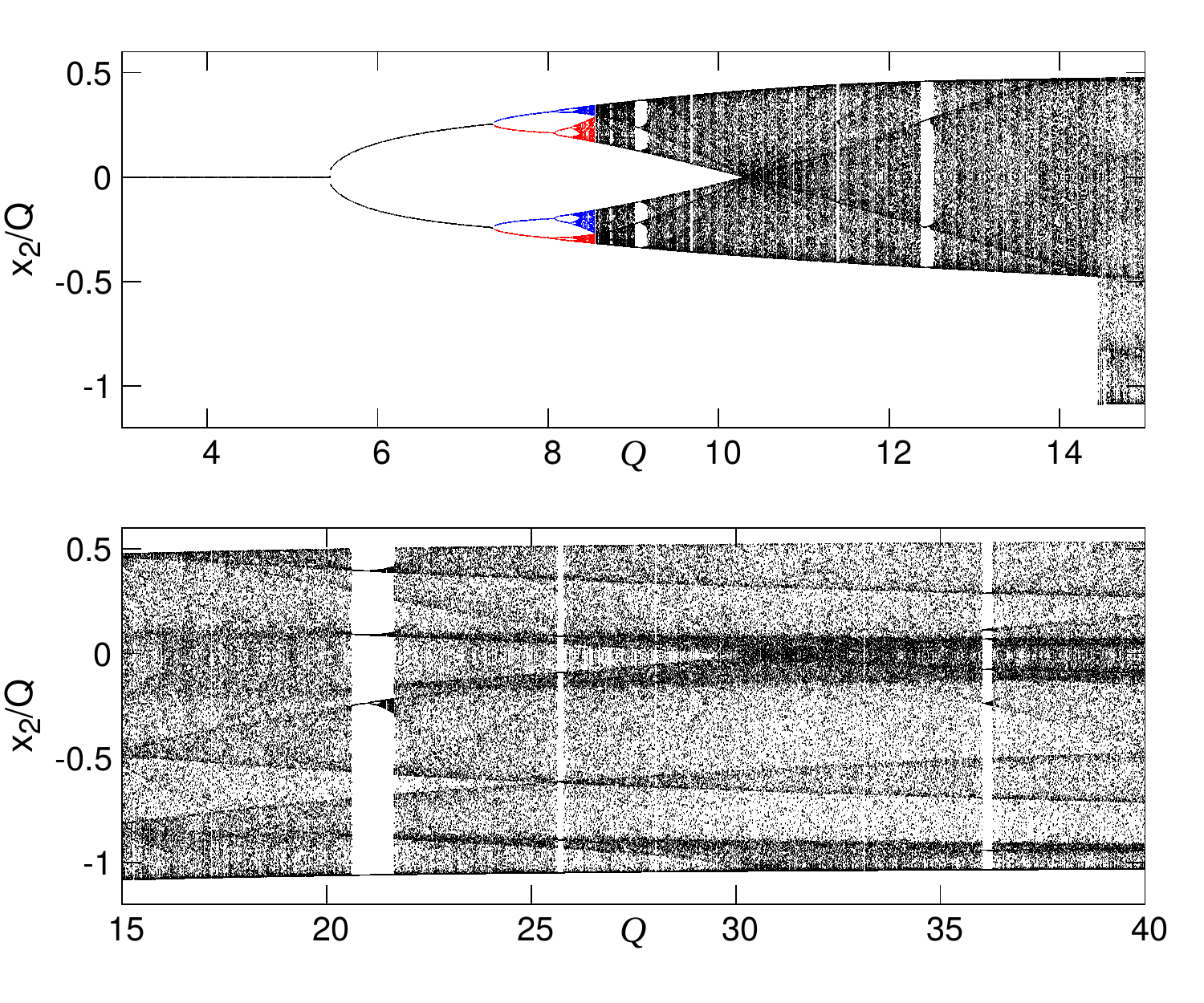}
\caption{Bifurcation diagram obtained at the Poincare section $x_1=0$, $\dot x_1>0$; depicted are values of the variable $x_2$ at these cross-sections.}
\label{fig:manbd}
\end{figure}

\subsection{Periodic dynamics and its transversal stability}

\label{sec:nesm}

Figure \ref{fig:manM1} shows the bifurcation diagram of the 
periodic orbit on the symmetric manifold M1, 
and its stability (including transversal one).
Figure \ref{fig:manbd} shows the bifurcation diagram of the full 6-dimensional system. 
It has been constructed by following the cycle on manifold M1 starting from 
small values of $Q$. This cycle is stable for $Q\lesssim 5.43$ (as can be seen from the plot of multipliers Fig.~\ref{fig:manM1}), at this value it experiences
period-doubling.  A next transition is at $Q\approx 7.36$, where through a pitchfork bifurcation, two mutually symmetric cycles arise. These cycles are shown in Fig.~\ref{fig:manbd} with two different colors (red and blue). Further increase of values of $Q$ leads to a transition to chaos. We will discuss this part of the diagram in the next section \ref{sec:nexp}, starting from large values of $Q$ and following decrease of these values.

\section{Numerical exploration of chaos in coupled heteroclinic cycles}
\label{sec:nexp}
\subsection{Bifurcation diagram}
For fixed values $\alpha,\beta$ \eqref{eq:spm}, the only bifurcation parameter is the strength of coupling $Q$. The diagram of the states
in dependence on $Q$ is presented in Fig.~\ref{fig:manbd}. Here
we show  values of the variable $x_2$
at the Poincar\'e surface of section $x_1=0,\dot x_1>0$. One can see that in
a large range of values of $Q$ the dynamics is chaotic, with some periodic windows present. 

\subsection{Symmetric chaotic attractor for weak coupling (large $Q$)}

\begin{figure}[!htb]
\centering
\includegraphics[width=0.85\textwidth]{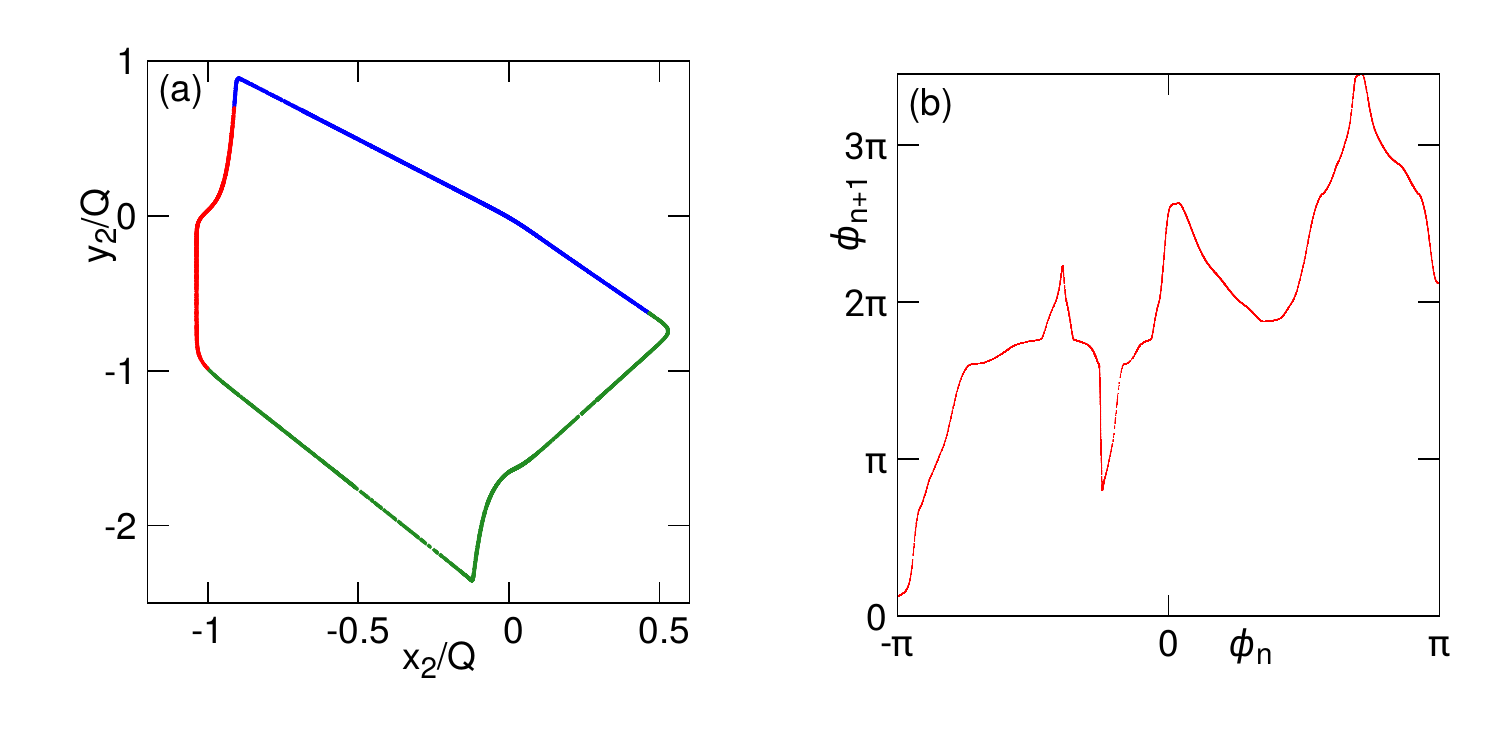}
\caption{Panel (a): Points at the Poincar\'e map projected on plane $(x_2,y_2)$.
Although the figure looks like a curve, in fact it is a dense set of points. Colors of these points indicate the closest invariant synchronous manifolds 
(blue: M1; green: M2; red: M3). Panel (b): a one-dimensional map constructed as described in text.
}
\label{fig:pm1}
\end{figure}

We start with chaotic states at weak coupling (large values of $Q$, bottom panel at Fig.~\ref{fig:manbd}). We take $Q=35$ and explore the Poincar\'e section $x_1=0$, 
$\dot x_1>0$ like in Fig.~\ref{fig:manbd}. At this section, we plot the values of variables 
$(x_2,y_2)$ in Fig.~\ref{fig:pm1}(a). This image looks like a closed curve, but in fact it is a set of points. This indicates that, with a good accuracy,
the system's dynamics is described by a one-dimensional map of a circle onto itself.
To construct this map, we rescale the variables and attribute to each
point in Fig.~\ref{fig:pm1}(a) an angle $\phi_n=\arctan(\eta_n/\xi_n)$,
where $\xi=(x_2+15)/56$ and $\eta=(y_2+20)/117$. Figure \ref{fig:pm1}(b)
shows the transformation $\phi_n\to \phi_{n+1}$. One can see that
it looks like a non-invertible circle map. This map has many
expanding regions, producing chaos. However, the map is continuous with
small regions where its derivative vanish; such regions give rise to periodic windows.

It is instructive to explore the role of symmetric manifolds M1,M2,M3 (cf. Eqs.~\eqref{eq:mf1},\eqref{eq:mf23}). We define the closeness to these manifolds
via the observables
\begin{equation}
\begin{gathered}
d_{M1}=(x_1-x_2)^2+(y_1-z_2)^2+(z_1-y_2)^2\;,\quad d_{M2}=(x_1-z_2)^2+(y_1-y_2)^2+(z_1-x_2)^2\;,
\\ d_{M3}=(x_1-y_2)^2+(y_1-x_2)^2+(z_1-z_2)^2\;,
\end{gathered}
\end{equation}
which vanish exactly on the manifolds.

\begin{figure}[!htb]
\centering
\includegraphics[width=0.8\textwidth]{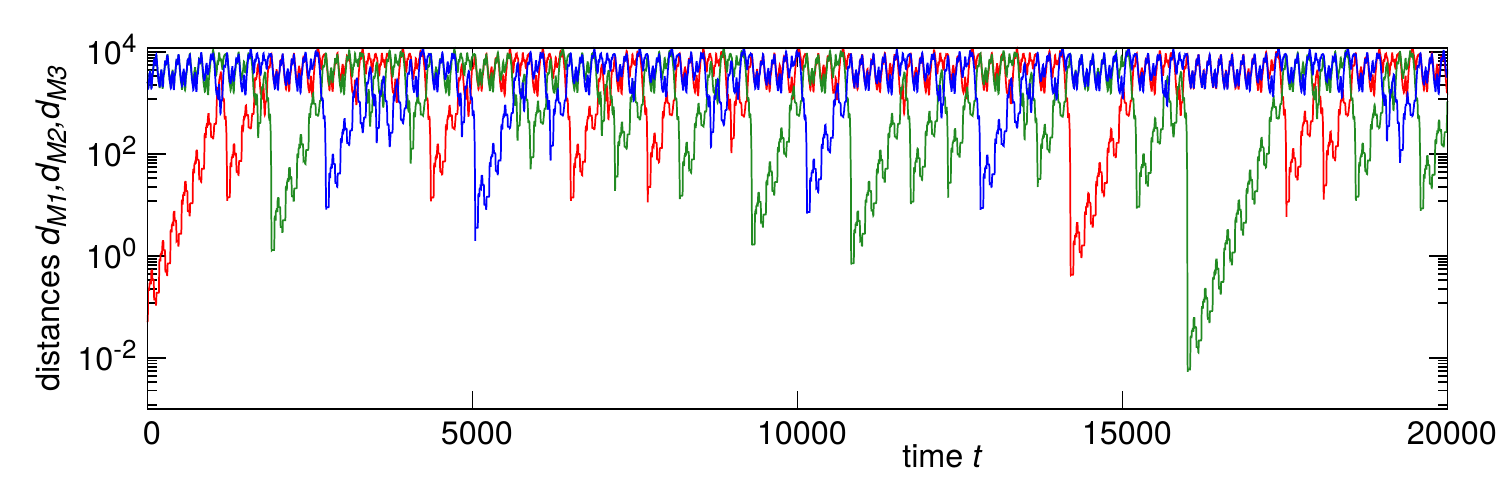}
\caption{Characterization of closeness of a trajectory to the submanifolds M1,M2,M3,
at $Q=35$. The lines show quantities $d_{M1}$ (blue); $d_{M2}$ (green); $d_{M3}$ (red).
}
\label{fig:tr-dm}
\end{figure}

The time evolution of these observables at $Q=35$ is shown in Fig.~\ref{fig:tr-dm}.
One can see that most of the time one of these variables is much smaller
than another ones (notice the logarithmic scale of these distances). This allows for a representation of the dynamics as an irregular sequence of nearly synchronous patches. In Fig.~\ref{fig:pm1}(a) we use the
same colors as in Fig.~\ref{fig:tr-dm}, to indicate positions on 
this Poincar\'e map, corresponding to the vicinities of different
synchronous manifolds.

\begin{figure}[!htb]
\centering
\includegraphics[width=0.8\textwidth]{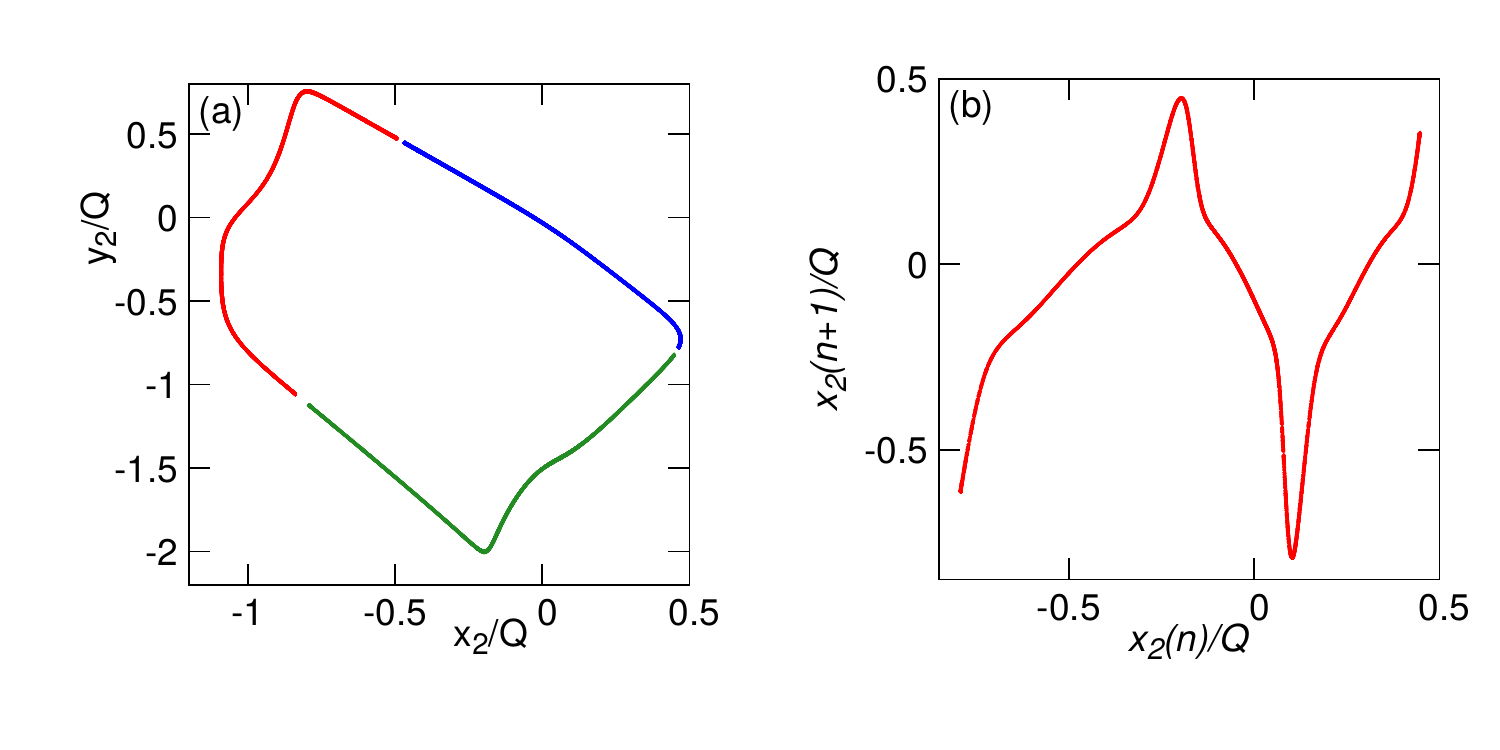}
\caption{Panel (a) Three Poincare maps with the same section for three symmetric attractors
at $Q=14$. Panel (b) A first return map $x_2(n)\to  x_2(n+1)$ for the green attractor in panel (a).
 This piece is chosen because its projection
on the variable $x_2$ is well-defined.}
\label{fig:pm2}
\end{figure}

\begin{figure}[!htb]
\centering
\includegraphics[width=0.7\textwidth]{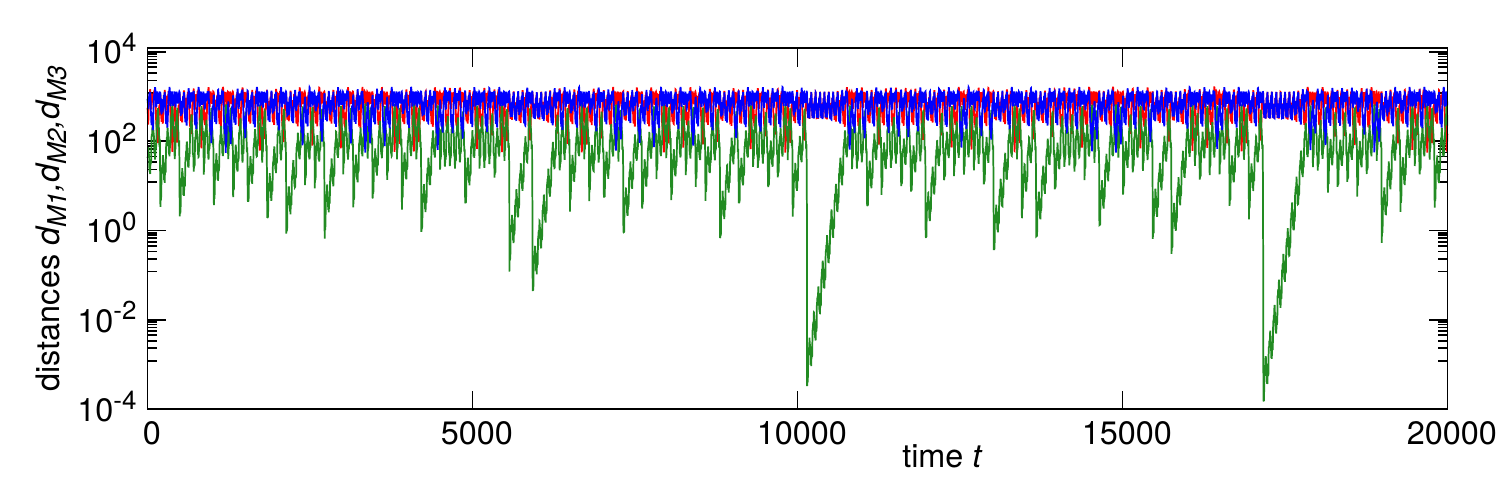}
\caption{Distances to the manifolds M1-M3 for the green attractor
of Fig.~\ref{fig:pm2} at $Q=14$. For other attractors, the picture is the same, with the corresponding rotation of colors.}
\label{fig:tr-dm2}
\end{figure}

\subsection{Symmetry breaking transitions} 

Here, we describe several symmetry-breaking transitions that occur as the coupling
becomes stronger (parameter $Q$ decreases). 

The first transition, at which the attractor of
Fig.~\ref{fig:pm1} breaks into three symmetric attractors, 
happens at $Q\approx 14.5$. This transition is clearly seen as a jump
in the bifurcation diagram (Fig.~\ref{fig:manbd}). 
The Poincar\'e map for $Q=14$ is shown in
Fig.~\ref{fig:pm2}. One can see three pieces, each correspond to one of the
three attractors. Because of the symmetry, we present the effective
one-dimensional map just for one of these attractors in panel (b).
One can see that it looks like a distorted $\sin$-map.

\begin{figure}[!htb]
\centering
\includegraphics[width=0.8\textwidth]{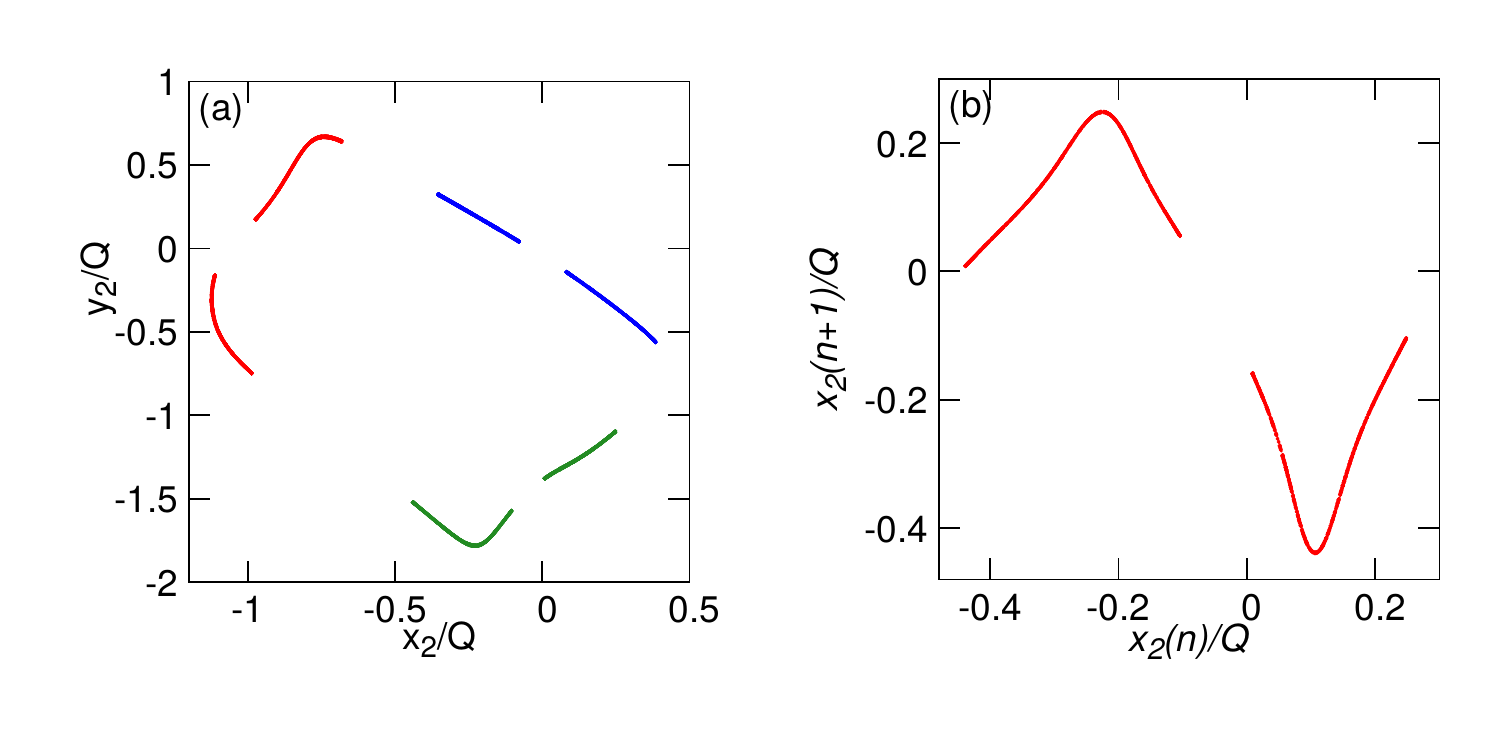}
\caption{Panel (a) Three Poincare maps with the same section for three symmetric attractors at $Q=9.5$. Panel (b) A first return 
map $x_2(n)\to x_2(n+1)$ for the green attractor in
panel (a). This piece is chosen because its projection
on the variable $x_2$ is well-defined.}
\label{fig:pm3}
\end{figure}

\begin{figure}[!htb]
\centering
\includegraphics[width=0.8\textwidth]{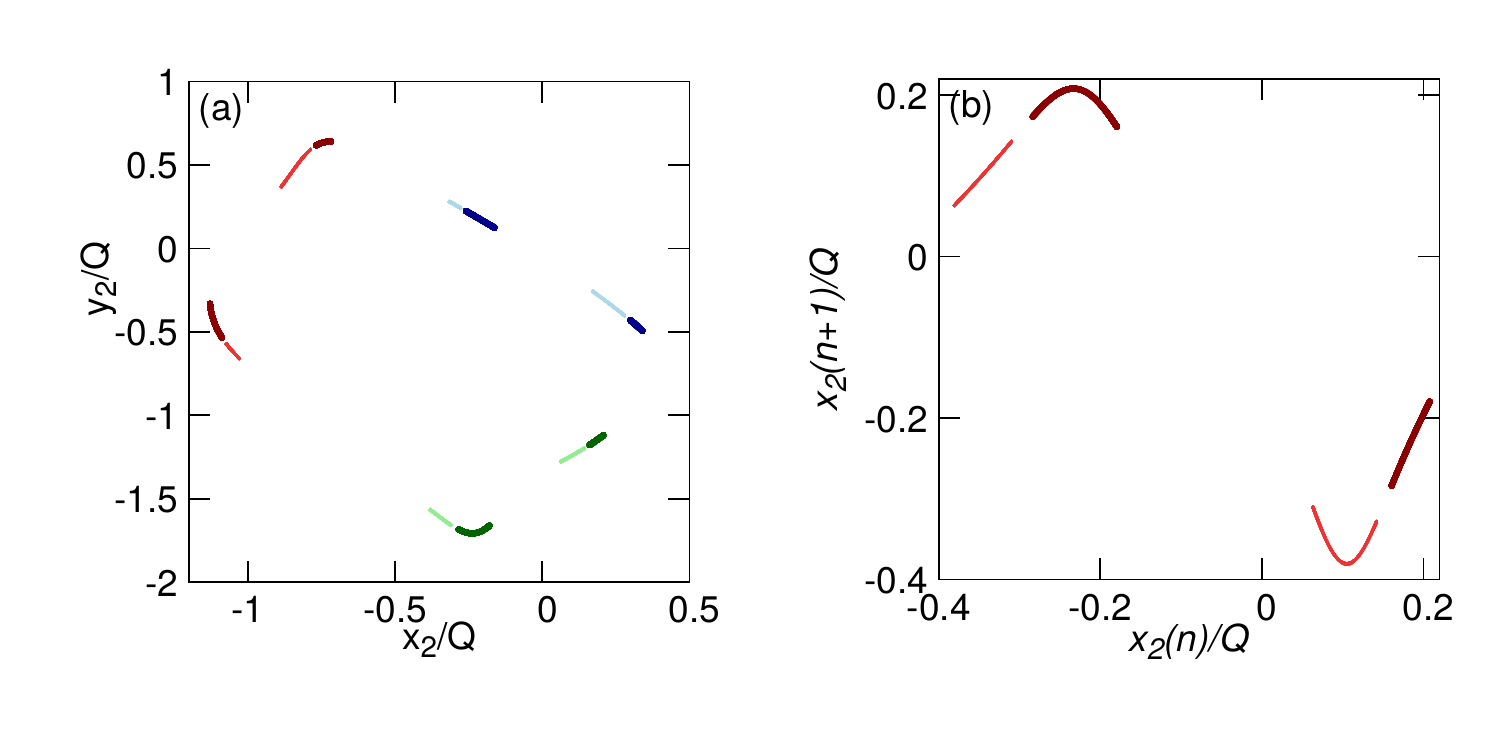}
\caption{Panel (a) Six Poincare maps with the 
same section for three symmetric attractors at $Q=8.5$. 
We use the same colors as in Fig.~\ref{fig:pm3}, but the two attractors
are shown with thin and bold markers and using a dark and a light version of the color.
Panel (b) A first return 
map $x_2(n)\to x_2(n+1)$ for the green attractors in
panel (a). Here again two attractors are shown with different size of the points
and with dark and light versions of the color.}
\label{fig:pm4}
\end{figure}

We show the time evolution of the distances to the submanifolds M1-M3 
for the green attractor of Fig.~\ref{fig:pm2}
in Fig ~\ref{fig:tr-dm2}. One can see that now the dynamics is persistently
close to the symmetric manifold M2.

At the next transition (at $Q\approx 10$), each of the symmetric attractors
splits into two period-two bands. The attractors and the corresponding
first-return map are shown in Fig.~\ref{fig:pm3}. 
At the next transition (at $Q\approx 8.55$) each of three attractors splits into two,
as demonstrated in Fig.~\ref{fig:pm4}. With further decrease of value of $Q$, these attractors undergo an inverse cascade of period-doublings. At $Q\approx 8.08$ stable period-2  asymmetric cycle at the end of this cascade
appears. At $Q\approx 7.36$, through an inverse pitchfork bifurcation, two asymmetric stable period-2 cycles merge into a stable symmetric period-2 cycle, as
described in Section \ref{sec:nesm} above.

\section{Piecewise-constant model of heteroclinic cycle}
\label{sec:pla}
In this section, we construct an approximate piecewise-constant (PC) model
of the heteroclinic cycle. It is based on its representation in variables 
\eqref{eq:xyzvar} within system \eqref{eq:tr}. The functions appearing in
 \eqref{eq:tr} resemble an expression for a Fermi distribution, which has a simple
 form in the zero-temperature limit  $T\to 0$
 \begin{equation}
    \frac{1}{1+e^{-x}}\quad\Leftrightarrow\quad  \frac{1}{1+e^{\epsilon/T}}=\begin{cases} 0&\epsilon>0\;,\\ 1 & \epsilon<0\;.
\end{cases}
\label{eq:fermi}
 \end{equation}
In our case, there is no small parameter $T$, 
but one can consider
the range of variations of the variable $x$ (which goes to large positive and large negative values) as a large parameter. Then, a smooth transition in
a region $x\approx 0$ can be approximated with a step function, 
similarly to the zero-temperature limit in \eqref{eq:fermi}. 

\begin{figure}[!htb]
\centering
\includegraphics[width=0.3\textwidth]{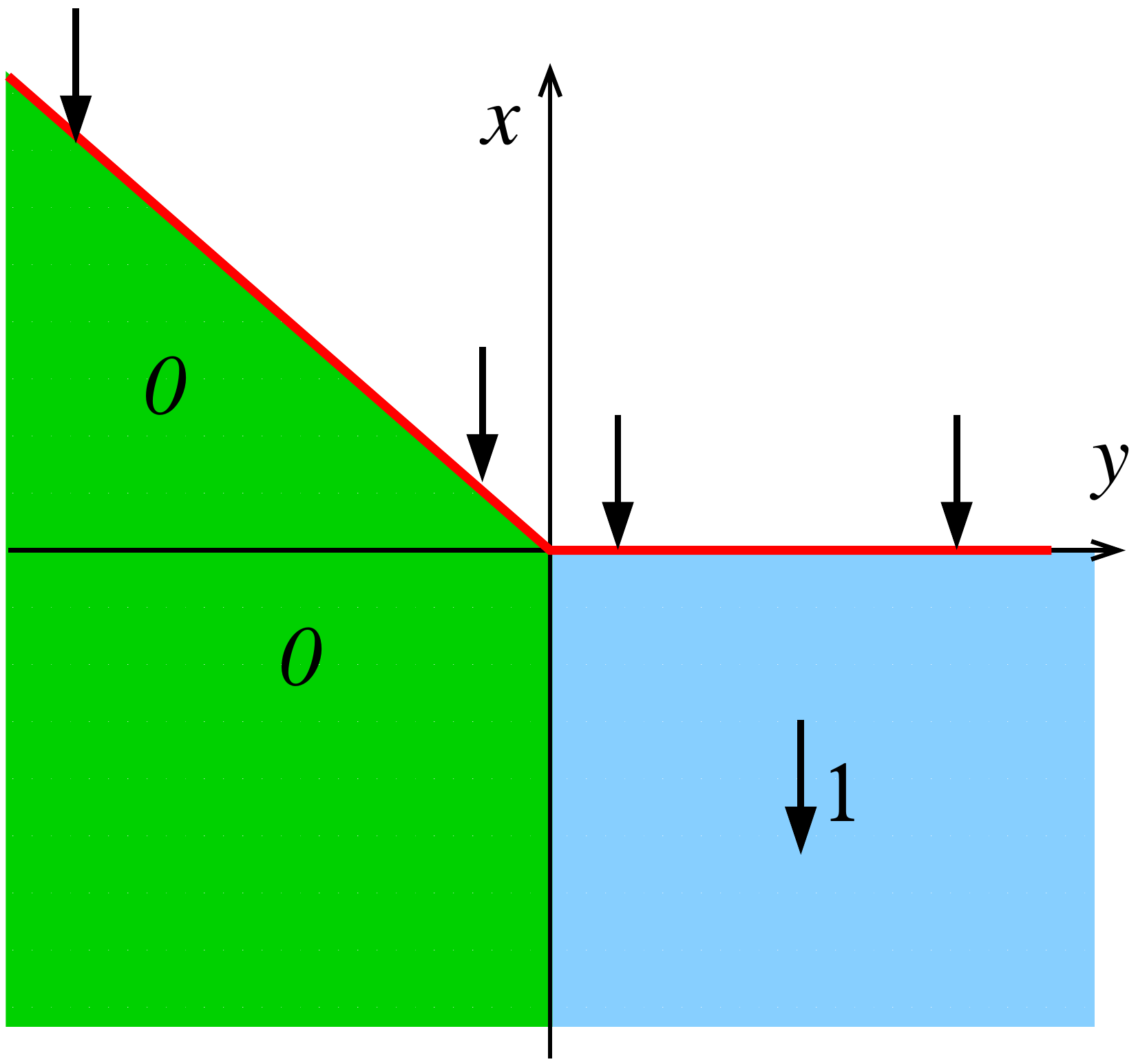}
\caption{Piecewise-constant representation of function $F(x,y)$,
contributing to the dynamics of variable $x$. The white area is a no-go domain where 
formally $F=\infty$. Arrows show the direction of the evolution of $x$ according to this term.
}
\label{fig:funF}
\end{figure}

Let us consider, as an example, the second term in the equation for $x$ in \eqref{eq:tr}.
We can approximate it according to the values of $(x,y)$ as follows:
\begin{equation}
\frac{1+e^x}{1+e^{-y}}\approx F(x,y)=\begin{cases}
\infty &\text{for } x>0,y>0 \text{ and } y<0,x>-y\;,\\
0 & \text{for } y<0,x<-y\;,\\
1 & \text{for } y>0,x<0\;.
\end{cases}
\label{eq:apprx}    
\end{equation}
The domain where the value of this term is $\infty$ is in fact a no-go area,
it is never entered during the dynamics. In the rest available area, this term takes values $1$ and $0$. We illustrate this in Fig.~\ref{fig:funF}. The available domain
$(y<0,x<-y)\cup (y>0,x<0)$ has two borders: $(x=-y; y<0)$ and $(x=0;y>0)$. The latter border
is not reachable, because there $\dot x<0$. However, at the former border the time derivative
of $x$ according to the term \eqref{eq:apprx} vanishes. Thus, a piece of trajectory
along this border $(x=-y; y<0)$ is possible, if the variable $y$ grows. 

\begin{table}[!htb]
\centering
\begin{tabular}{|p{0.2\textwidth}|p{0.14\textwidth}|p{0.14\textwidth}|p{0.1\textwidth}|}
\hline
Regime & Conditions & Equations & Next regime\\ \hline
 A &$x>0$\newline  $|y|>|x|$&
$\dot x =1$\newline $\dot y=1-\beta$\newline $\dot z=1-\alpha$& $\to$ Ab\\ \hline
 Ab (boundary of A) &$x>0$\newline $|y|=|x|$ & $x=-y$\newline $\dot y=1-\beta$
\newline $\dot z=1-\alpha$& $\to$ B\\
\hline
 B & $y>0$\newline $|z|>|y|$&
$\dot x =1-\alpha$\newline $\dot y=1$\newline $\dot z=1-\beta$ & $\to$ Bb\\ \hline
 Bb (boundary of B) &$y>0$\newline $|z|=|y|$ & $\dot x=1-\alpha$\newline 
$ y=-z$\newline $\dot z=1-\beta$& $\to$ C\\
\hline
 C & $z>0$\newline $|x|>|z|$&
$\dot x =1-\beta$\newline $\dot y=1-\alpha$\newline $\dot z=1$ & $\to$ Cb\\ \hline
Cb (boundary of C) &$z>0$\newline $|x|=|z|$ & $\dot x=1-\beta$\newline $\dot y=1-\alpha$\newline $ z=-x$& $\to$ A\\
\hline
\end{tabular}
\caption{Piecewise-constant dynamics, approximating the heteroclinic cycle
in Eqs.~\eqref{eq:tr} according to \eqref{eq:apprx}. }
\label{tab:t1}
\end{table}

\begin{figure}[!htb]
\centering
\includegraphics[width=0.8\textwidth]{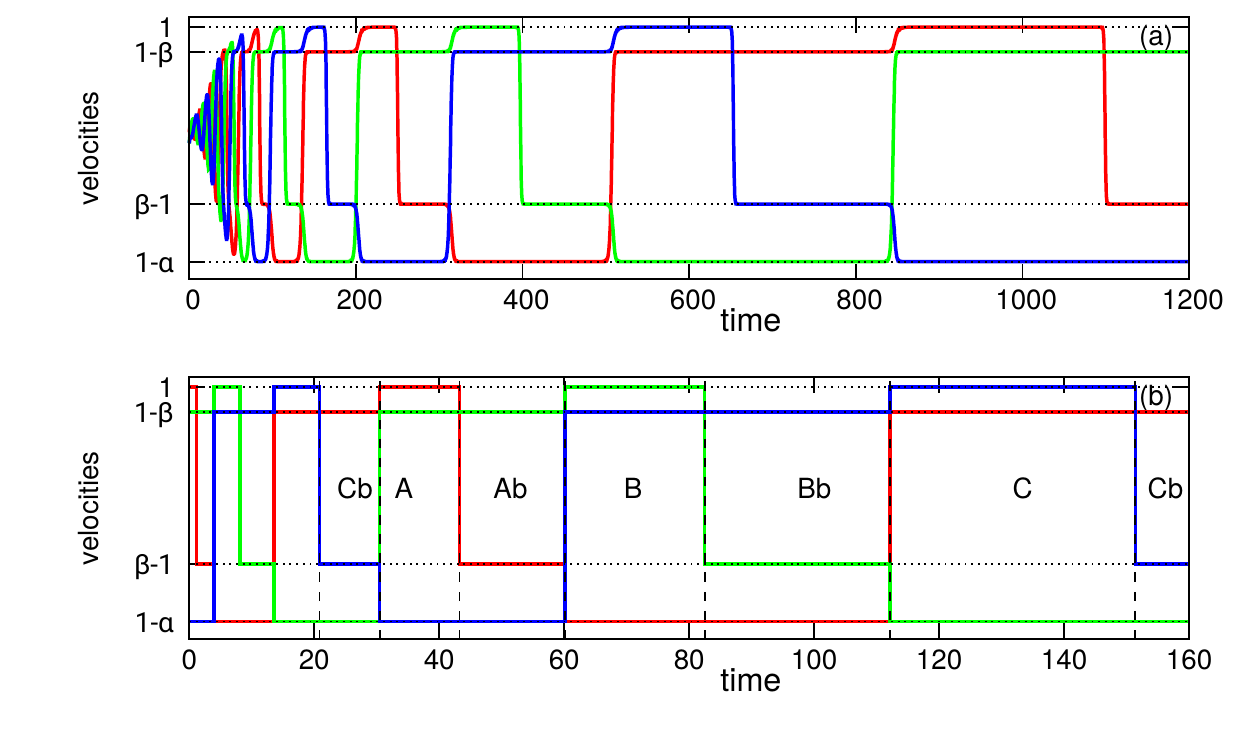}
\caption{Heteroclinic cycle in Eqs.~\eqref{eq:tr} (panel (a)) and in the 
PC representation Table \ref{tab:t1} (panel (b)). Depicted are the time derivatives $\dot x$ (red), $\dot y$ (green), and $\dot z$ (blue)
as functions of time along a trajectory. One can see a good correspondence, except
for an initial transient in panel (a), during which a trajectory approaches
the heteroclinic cycle. Dashed lines in panel (b) show the switchings between the piecewise-constant patches; these patches are shown with the same coding as in Table \ref{tab:t1}.
}
\label{fig:2c}
\end{figure}

The approximation \eqref{eq:apprx} can be applied to all the terms on the r.h.s. of \eqref{eq:tr}. It can be summarized in a sequence of 
possible constant-velocity pieces listed in Table \ref{tab:t1} (there, for brevity,
we list only pieces along a trajectory close to the heteroclinic cycle for $\alpha>1$, $\beta<1$). 
Matching these pieces $A\to Ab\to B\to Bb\to C\to Cb\to A\to \ldots$ yields a full trajectory approaching
 the heteroclinic cycle (see Fig.~\ref{fig:2c}).

\section{Piecewise-constant model of coupled heteroclinic cycles}
\label{sec:pc}
\subsection{Effect of coupling in the PC approximation}

We now extend the analysis of Section~\ref{sec:pla} above to the case
of coupled heteroclinic cycles \eqref{eq:trc}. Let us consider the equation for $x_1$,
where the additional term is 
\[
C(x_1,x_2)=e^{-Q}\frac{(1+e^{-x_1})(1+e^{x_1})}{1+e^{-x_2}}-(1+e^{x_1})\;.
\]
Similarly to \eqref{eq:apprx} we obtain in the PC approximation
\begin{equation}
C(x_1,x_2)=\begin{cases} -\infty & \text{for } (x_2>0,x_1>x_2+Q) \text{ and }
(x_2<0, x_1>Q)\;,\\
0 & \text{for } (x_2>0,Q<x_1<x_2+Q) \text{ and }
(x_2<0, -Q+x_2<x_1<Q)\;,\\
\infty & \text{for } (x_2>0,x_1<-Q) \text{ and }
(x_2<0, x_1<-Q+x_2)\;.
\end{cases}
\label{eq:dom1}
\end{equation}

\begin{figure}[!htb]
\centering
(a) \includegraphics[width=0.35\textwidth]{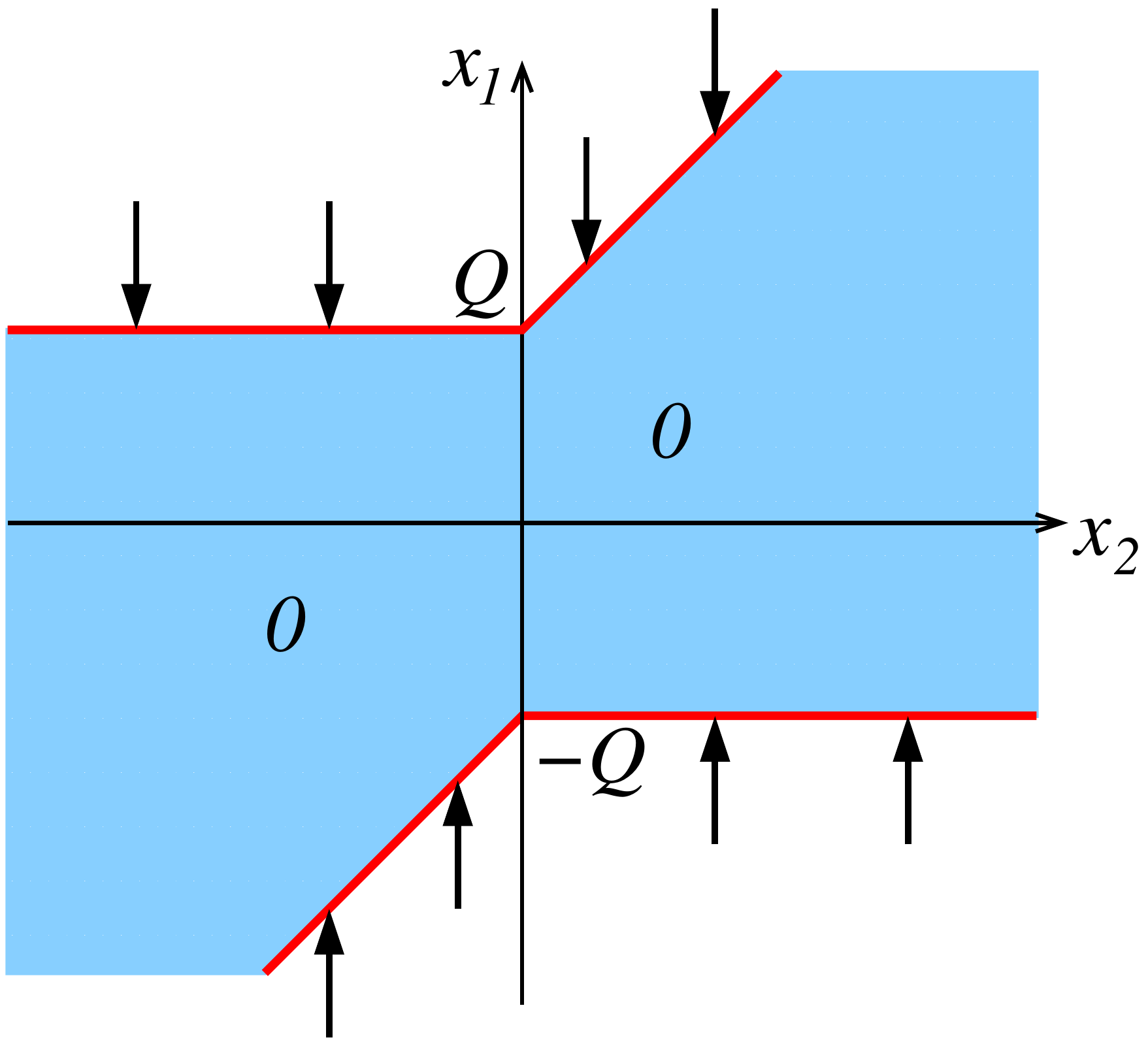}\qquad
(b) \includegraphics[width=0.35\textwidth]{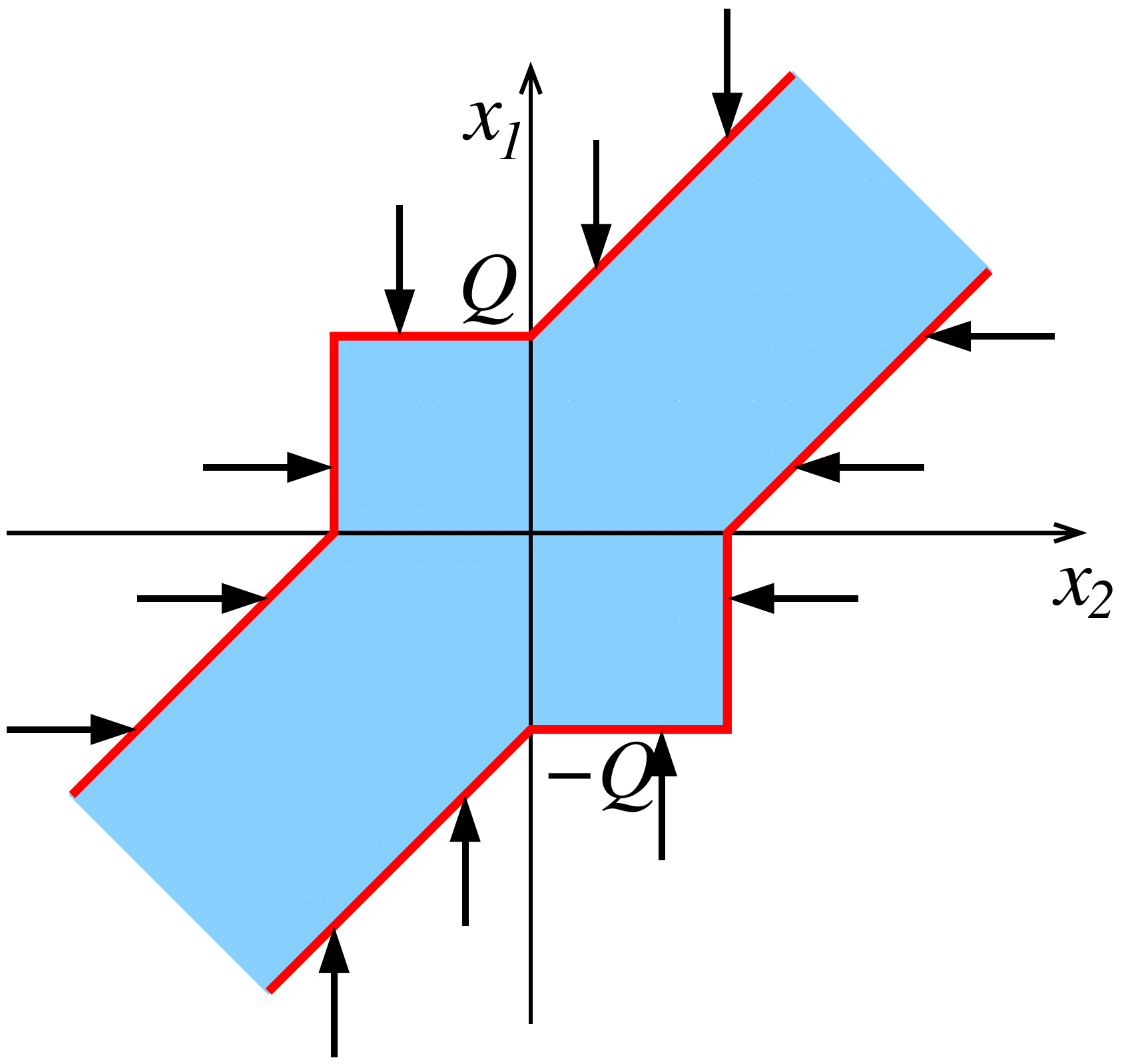}
\caption{Panel (a): Accessible domain for $x_1$ on plane $x_1,x_2$ according 
to Eqs.~\eqref{eq:dom1}. Panel (b): Full accessible for $x_1$ and $x_2$.} 
\label{fig:domc}
\end{figure}

The available domain where $C=0$ is depicted in Fig.~\ref{fig:domc}(a).
If we combine this domain with the symmetric one resulting from the equation for
$x_2$, we obtain the effect of coupling as an accessible ``corridor'' in Fig.~\ref{fig:domc}(b).
Similar corridors restrict the dynamics of variables $(y_1,y_2)$ and $(z_1,z_2)$.

\subsection{Scaling}
Let us consider scaling of functions $F(x,y)$ (Eq.~\eqref{eq:apprx})
and $C(x_1,x_2)$ (Eq.~\eqref{eq:dom1}). The function $F(x,y)$ is scale-free.
The coupling function  $C(x_1,x_2)$ can be written as a universal
function of the normalized variables $C(x_1,x_2)=c\left(\frac{x_1}{Q},\frac{x_2}{Q}\right)$. Thus, in the PC
approximation the equations for the variables $x_1,y_1,z_1,x_2,y_2,z_2$
can be renormalized 
\begin{equation}
x_k=Q \tilde{x}_k,\; y_k=Q\tilde{y}_k,\;,z_k=Q \tilde{z}_k,\qquad t=Q \tilde{t},\qquad (k=1,2)\;.
\label{eq:sc}
\end{equation}
In the new equations, parameter $Q$ is replaced by one, and the only remaining
parameters are $\alpha,\beta$. This is the reason why in the figures above,
we have depicted variables $x,y,z$ scaled by factor $Q$.

The scaling property of the PC model means that 
there is no ``bifurcation diagram'' in dependence 
on the coupling strength $Q$ in this approximation. The dynamics for fixed $(\alpha,\beta)$ is universal (there can be still multistability, i.e. dependence on initial conditions). This is in a contradistinction to observation in the original system, where the bifurcation diagram is nontrivial (Fig.~\ref{fig:manbd}). However,
the part of Fig.~\ref{fig:manbd} at large $Q$ demonstrates predominantly chaos, and
the maximal and minimal values of variable $x_2$ scale $x_{max,min}\sim Q$
as the scaling \eqref{eq:sc} suggests. According to \eqref{eq:sc},
also the characteristic timescale (in particular, the period of a periodic orbit) scales $\sim Q$.  Below in this section, we use the PC approximation with $Q=1$.

\subsection{Synchronous cycle}

\begin{figure}[!htb]
\centering
\includegraphics[width=0.8\textwidth]{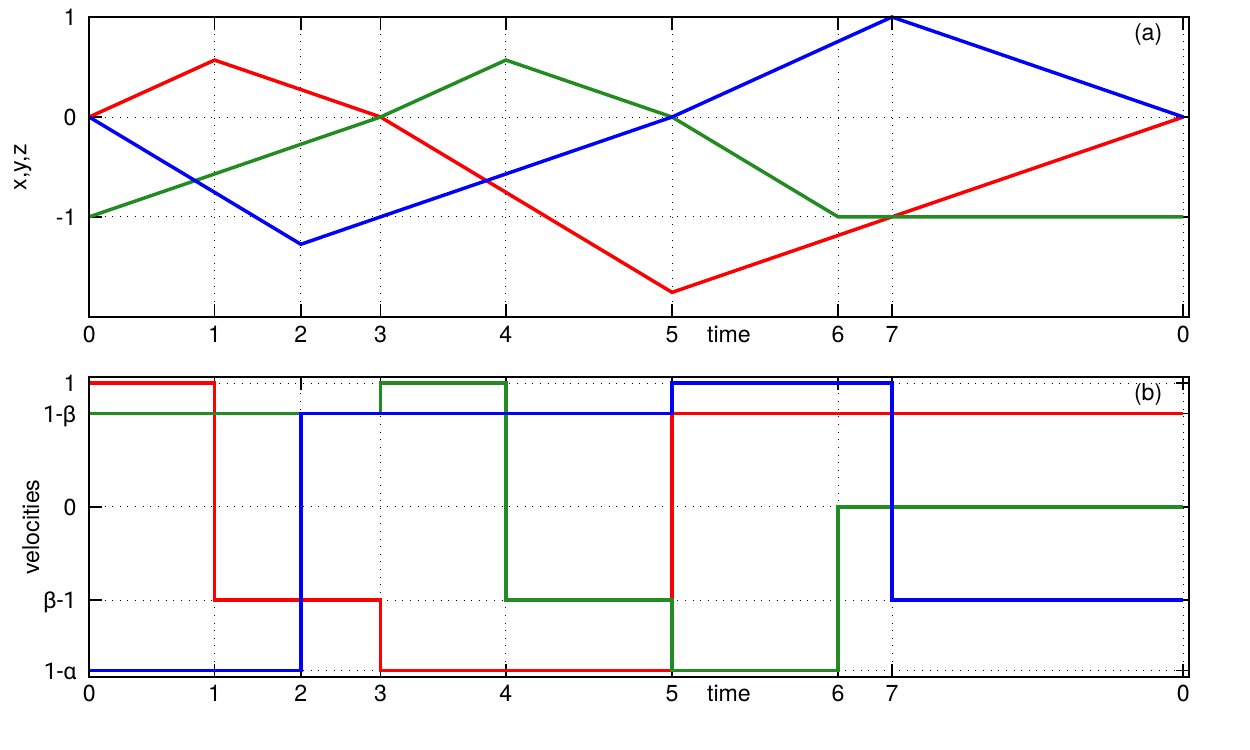}
\caption{Periodic cycle in the PC approximation. Panel (a): variables $x(t),y(t),z(t)$. Panel (b): their derivatives $\dot x(t),\dot y(t),\dot z(t)$.
Markers of the time axis indicate the instants of
switches of the dynamics.}
\label{fig:m1_tr}
\end{figure}

\begin{figure}[!htb]
\centering
\includegraphics[width=0.3\textwidth]{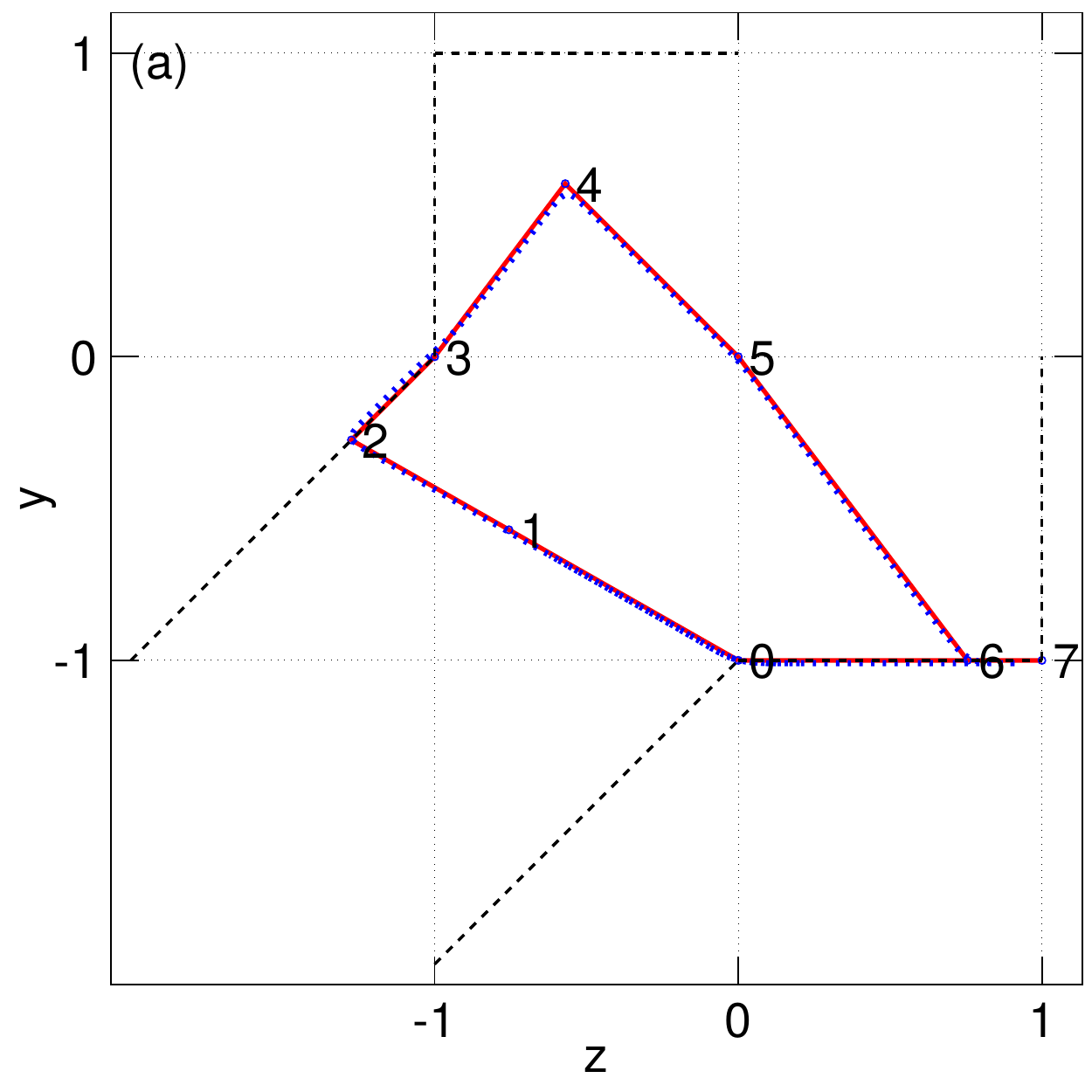}\hfill
\includegraphics[width=0.3\textwidth]{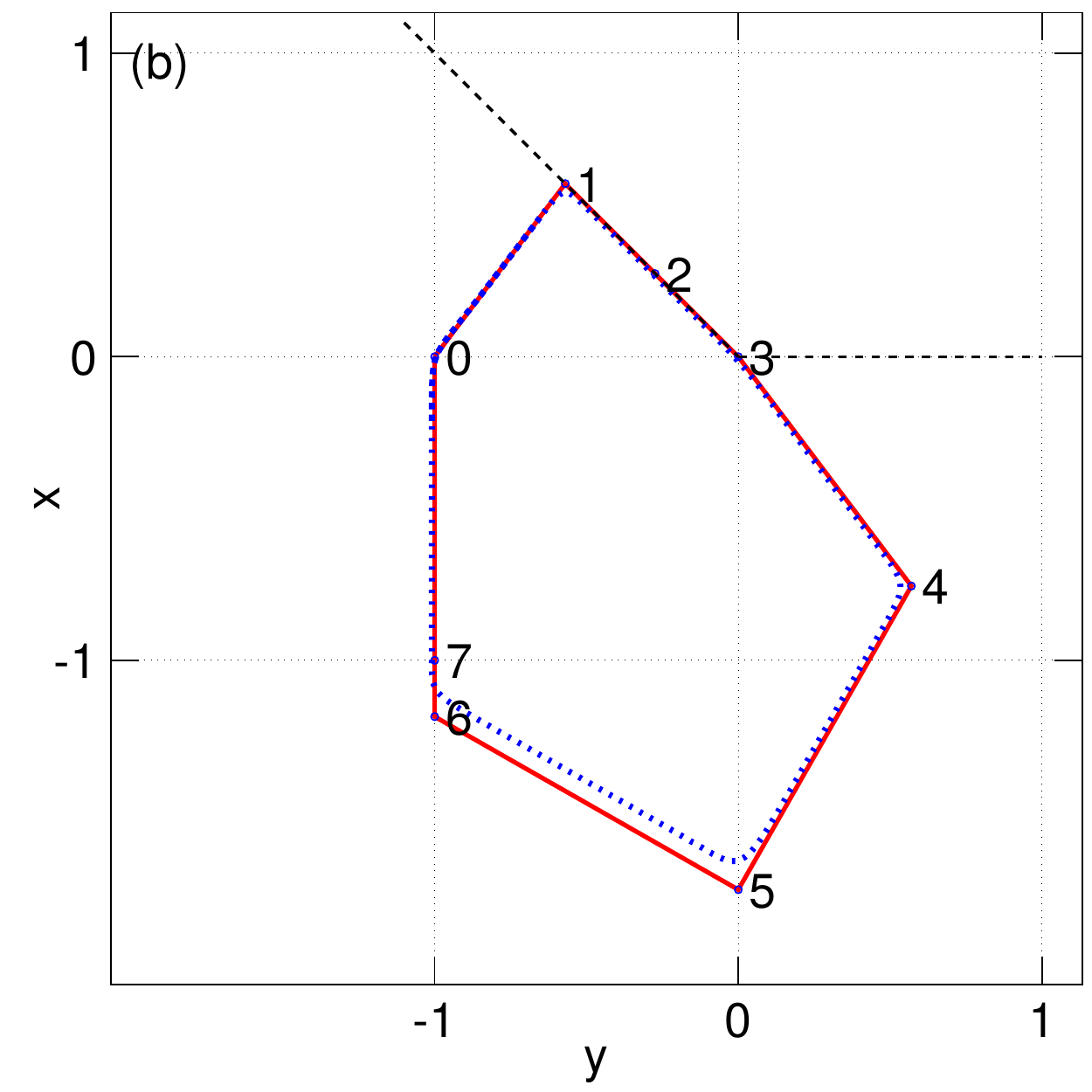}\hfill
\includegraphics[width=0.3\textwidth]{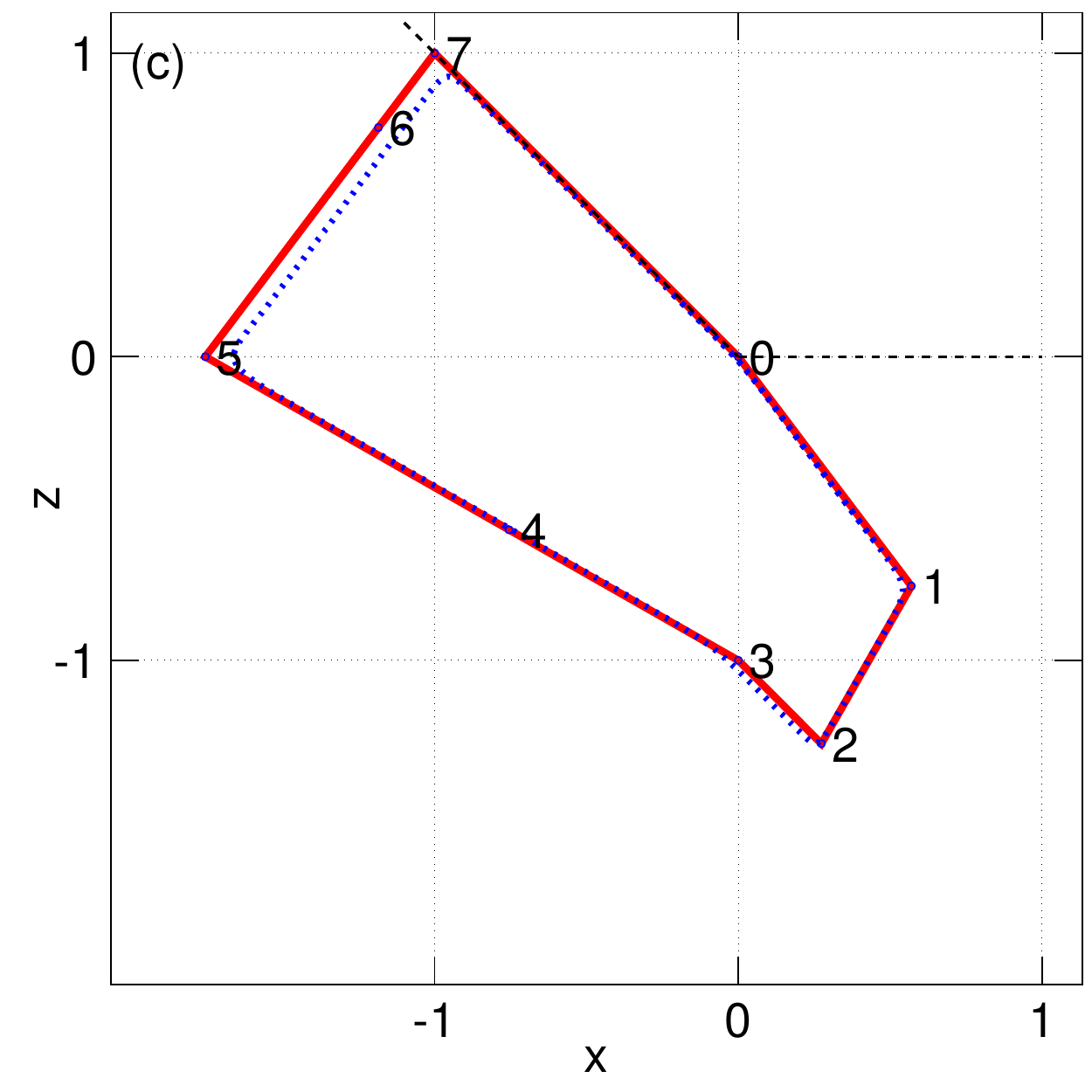}
\caption{Projections of the trajectory Fig.~\ref{fig:m1_tr}
on phase planes $(x,y),(y,z),(z,x)$ (red line). The numbers indicate the same 
instants of switches as in Fig.~\ref{fig:m1_tr}. The black dashed lines shows
borders of admissible regions according to Fig.~\ref{fig:funF} (panels (b),(c))
and to Fig.~\ref{fig:domc}(b) (panel (a)). The blue dotted
line shows the rescaled cycle for $Q=30$ in the full Eqs.~\eqref{eq:m1}.}
\label{fig:m1_pp}
\end{figure}

As the first example of application of the PC
approximation, we consider a symmetric cycle on the manifold M1 (cf.
Section \ref{sec:ism}). This cycle is described by Eqs.~\eqref{eq:m1}, which in variables $(x,y,z)$ read
\begin{equation}
\begin{aligned}
\dot x&=1-\alpha \frac{1+e^x}{1+e^{-y}}-\beta \frac{1+e^x}{1+e^{-z}}\;,\\
\dot y&=1-\alpha \frac{1+e^y}{1+e^{-z}}-\beta \frac{1+e^y}{1+e^{-x}}
+e^{-Q}\left[\frac{(1+e^{-y})(1+e^y)}{1+e^{-z}}-(1+e^y)\right]\;,\\
\dot z&=1-\alpha \frac{1+e^z}{1+e^{-x}}-\beta \frac{1+e^z}{1+e^{-y}}
+e^{-Q}\left[\frac{(1+e^{-z})(1+e^z)}{1+e^{-y}}-(1+e^z)\right]\;.
\end{aligned}
\label{eq:M1}
\end{equation}

The PC approximation of Eqs.~\eqref{eq:M1} is formulated as
\begin{equation}
\begin{aligned}
\dot x &=1-\alpha F(x,y) -\beta F(x,z)\;,\\
\dot y&=1-\alpha F(y,z)-\beta F(y,x)+C(y,z)\;,\\
\dot z&=1-\alpha F(z,x)-\beta F(z,y)+C(z,y)\;.
\end{aligned}
\label{eq:csfM1}
\end{equation}
The solution of these equations trivially scales
with parameter $Q$, but its form 
depends on parameters $\alpha,\beta$. For the values
of these parameters explored above (see Eq.~\eqref{eq:spm}), the cycle
is presented in Figs.~\ref{fig:m1_tr},\ref{fig:m1_pp}.
Points $0,\ldots,7$ indicate instants at which the 
dynamics changes; altogether the cycle consists of 8 pieces.
As Fig.~\ref{fig:m1_pp} shows, coupling term $\sim C$
is ``working'' only at stages $6\to 7 \to 0$ and $2\to 3$;
at other stages it is ``switched off''. Stages $1\to 2 \to 3$ and
$7\to 0$ follow the borders of the affordable domains according to function $F$. In the same figures we show the cycle in the full Eqs.~\eqref{eq:m1}
for $Q=30$ with blue dots; this gives an impression of accuracy
of the PC approximation.


\section{Chaotic and periodic regimes in 
the PC 
approximation.}
\label{sec:plach}

\begin{figure}[!htb]
\centering
\includegraphics[width=0.7\textwidth]{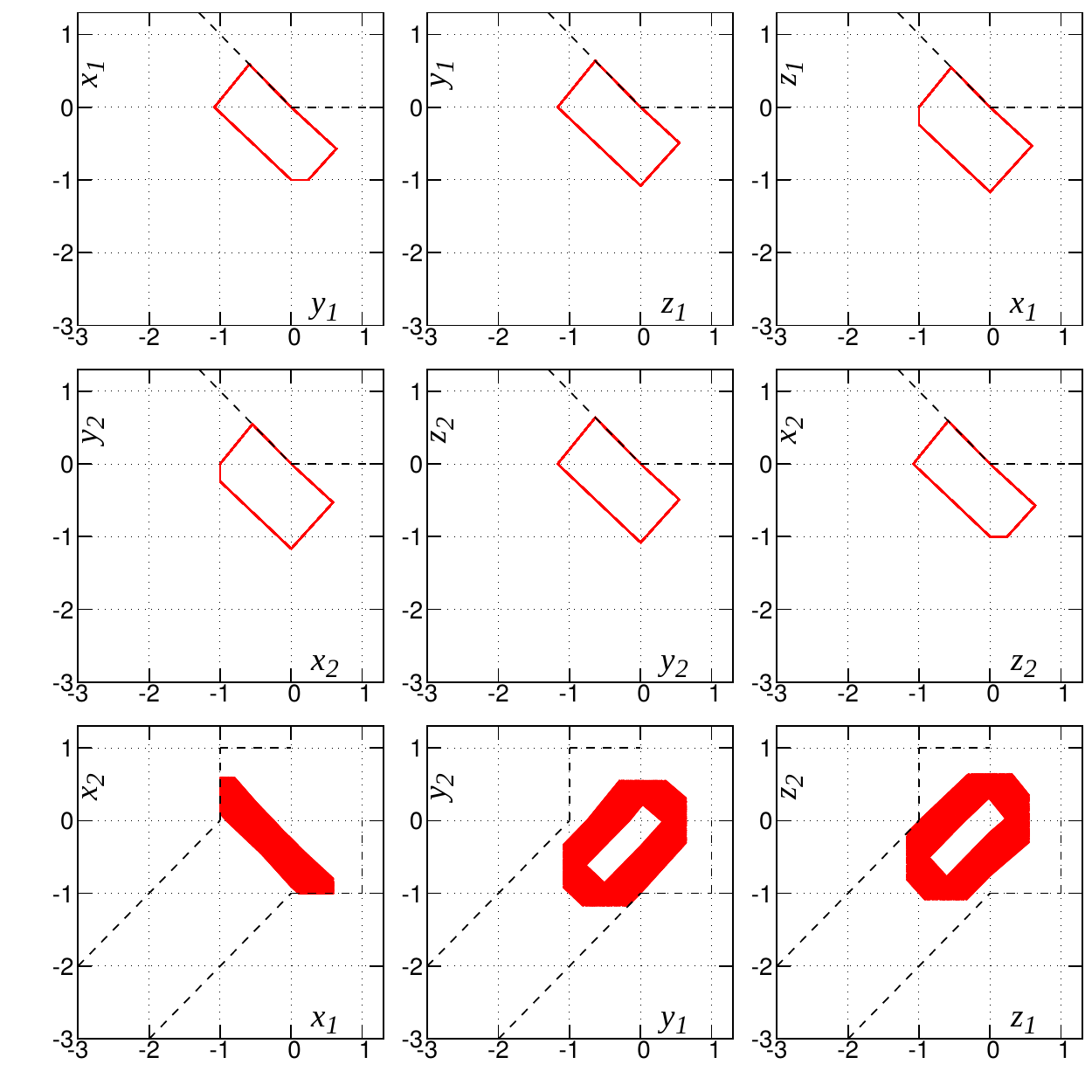}
\caption{Projections of many periodic orbits of type Aa at $\beta=1/6,\alpha=1.9$ on different subplanes. Dashed lines show available domains according to Figs.~\ref{fig:funF},\ref{fig:domc}.}
\label{fig:poa190}
\end{figure}

\begin{figure}[!htb]
\centering
\includegraphics[width=0.7\textwidth]{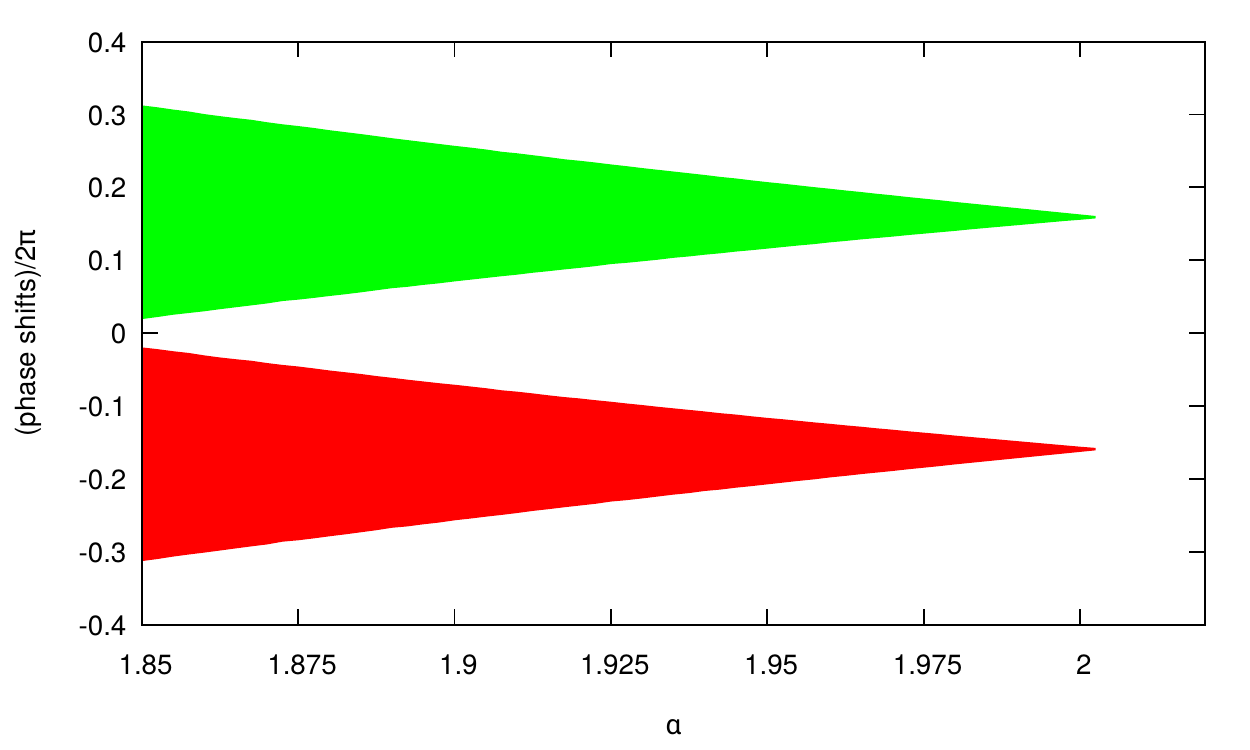}
\caption{The range of possible phase shifts $\Delta\varphi_y$ (red domain)
and $\Delta\varphi_z$ (green domain) between two subsystems for periodic orbits of type Aa. The phase shifts were calculated as follows. 
For a periodic orbit of period $T$, we calculated the times $t_{x_1,y_1,z_1}$ and $t_{x_2,y_2,z_2}$, at which the corresponding variable attains 
a maximum (there is one such point on the trajectory). After that, a phase shift in the pairs of variables was calculated
according to 
$
\Delta\varphi_x=2\pi(t_{x_2}-t_{x_1})/T$, $\Delta\varphi_y=2\pi (t_{y_2}-t_{y_1})/T$, $\Delta\varphi_z=2\pi (t_{z_2}-t_{z_1})/T$.
}
\label{fig:phsh}
\end{figure}

In this section we report on stable periodic and chaotic regimes in the PC model
of coupled heteroclinic cycles
\begin{equation}
\begin{aligned}
\dot x_1 &=1-\alpha F(x_1,y_1) -\beta F(x_1,z_1)+C(x_1,x_2)\;,\\
\dot y_1&=1-\alpha F(y_1,z_1)-\beta F(y_1,x_1)+C(y_1,y_2)\;,\\
\dot z_1&=1-\alpha F(z_1,x_1)-\beta F(z_1,y_1)+C(z_1,z_2)\;,\\
\dot x_2 &=1-\beta F(x_1,y_1) -\alpha F(x_1,z_1)+C(x_2,x_1)\;,\\
\dot y_2&=1-\beta F(y_1,z_1)-\alpha F(y_1,x_1)+C(y_2,y_1)\;,\\
\dot z_2&=1-\beta F(z_1,x_1)-\alpha F(z_1,y_1)+C(z_2,z_1)\;.
\end{aligned}
\label{eq:cPC}
\end{equation}
As mentioned above, the only relevant parameters are $\alpha$ and $\beta$. Below we 
fix $\beta=1/6$ and consider the range of other parameter $1.85\leq \alpha \leq 2.85$. The lower boundary is slightly larger than the value $\alpha_c=2-\beta=1.833$
at which the heteroclinic cycle in a single system appears. We observed 2 types of stable periodic orbits and two types of chaos.

\subsection{Stable periodic orbits of type A}
These periodic orbits build 3 families, which are transformed to each other
by the renaming transformation \eqref{eq:renam}, we denote these families as Aa, Ab, Ac. A remarkable property of these orbits is that the phase shift between two
periodic trajectories in systems 1 and 2 is to a large extent arbitrary. We illustrate this in Fig.~\ref{fig:poa190}. Here we overlap  160 periodic orbits of this type, appearing at different random initial conditions. One can see
that the trajectories projected on the planes of variables of subsystem 1 and of subsystem 2 coincide. However, projections on the planes showing variables of both subsystems do not coincide, because the phase shifts are different.

Periodic orbits of type A exist for small values of $\alpha\lesssim 2.0025$. In Fig.~\ref{fig:phsh} we show the range of possible phase shifts between the two subsystems.  

We attribute the existence of synchronous periodic orbits 
with phase shifts in some interval
to the property of existence of dead zones in coupling, recently
introduced in \cite{ashwin2019state,ashwin2021dead}. Indeed, in the PC model
coupling between the systems 1 and 2 is only at the borders of the corridors depicted in Fig.~\ref{fig:domc}(b); inside these corridors the systems do not interact. This is exactly the property of dead zones in the coupling terms, which, as has been shown in \cite{ashwin2019state,ashwin2021dead}, leads to an effective coupling function with
neutral intervals. Thus, phase shifts in some range become possible.

\begin{figure}[!htb]
\centering
\includegraphics[width=0.7\textwidth]{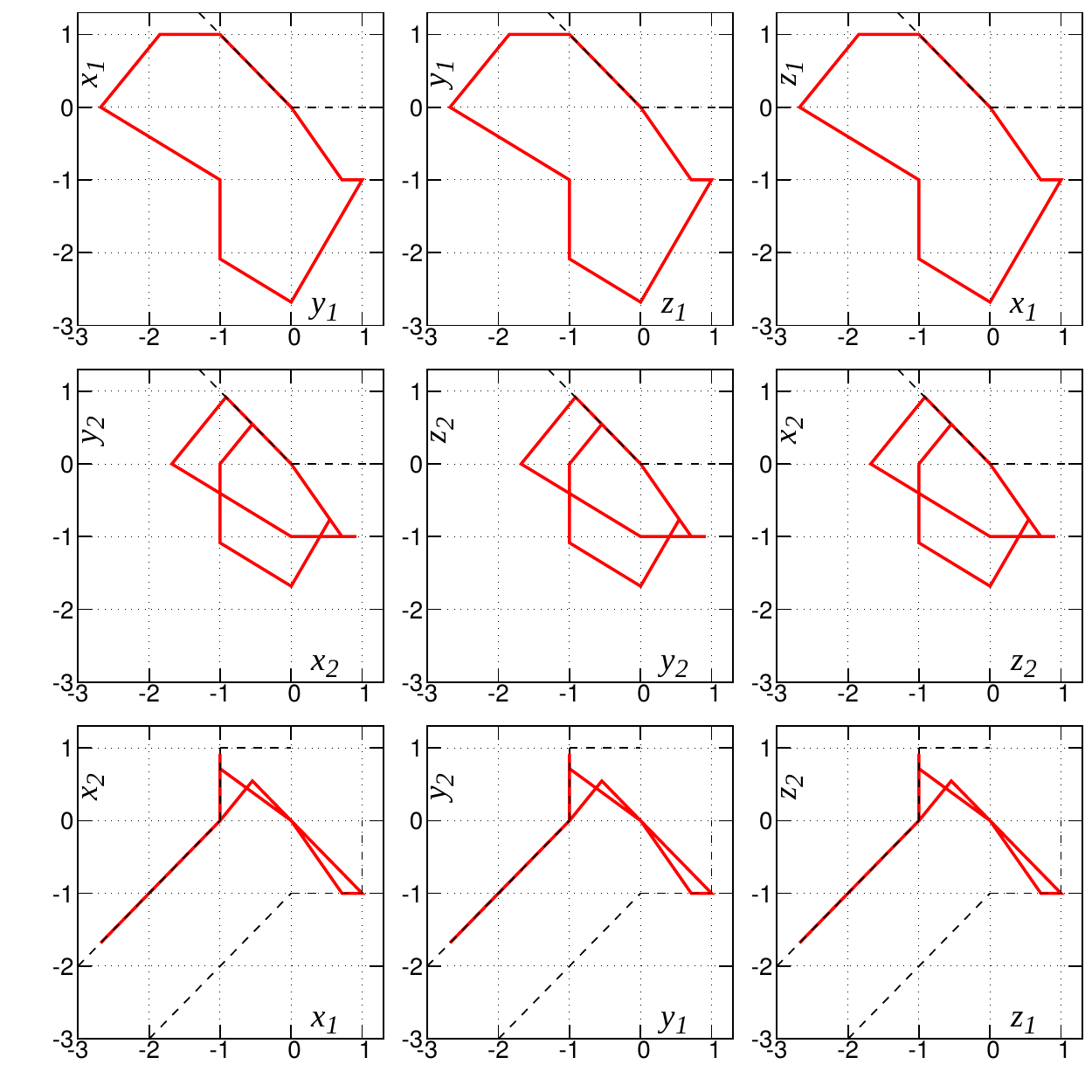}
\caption{Projection of a cycle of type B on different subplanes, for  $\beta=1/6,\alpha=2.4$. Dashed lines show available domains according to Figs.~\ref{fig:funF},\ref{fig:domc}.}
\label{fig:trB}
\end{figure}

\subsection{Stable periodic orbits of type B}
Stable periodic orbit of type B exists in the whole range of explored values of the parameter $\alpha$. This orbit is symmetric with respect to renamings, 
but asymmetric with respect to transformation $1\leftrightarrow 2$. 
We show projections of this trajectory on different subplanes
in Fig.~\ref{fig:trB}. This orbit can be characterized as a 2:1 synchronous one,
because here in system 1 there is one maximum and one minimum per period, while in system 2 there are two maxima and two minima. Correspondingly, the amplitude
in system 2 is smaller.

\subsection{Chaos of type A}

\begin{figure}[!htb]
\centering
\includegraphics[width=0.49\textwidth]{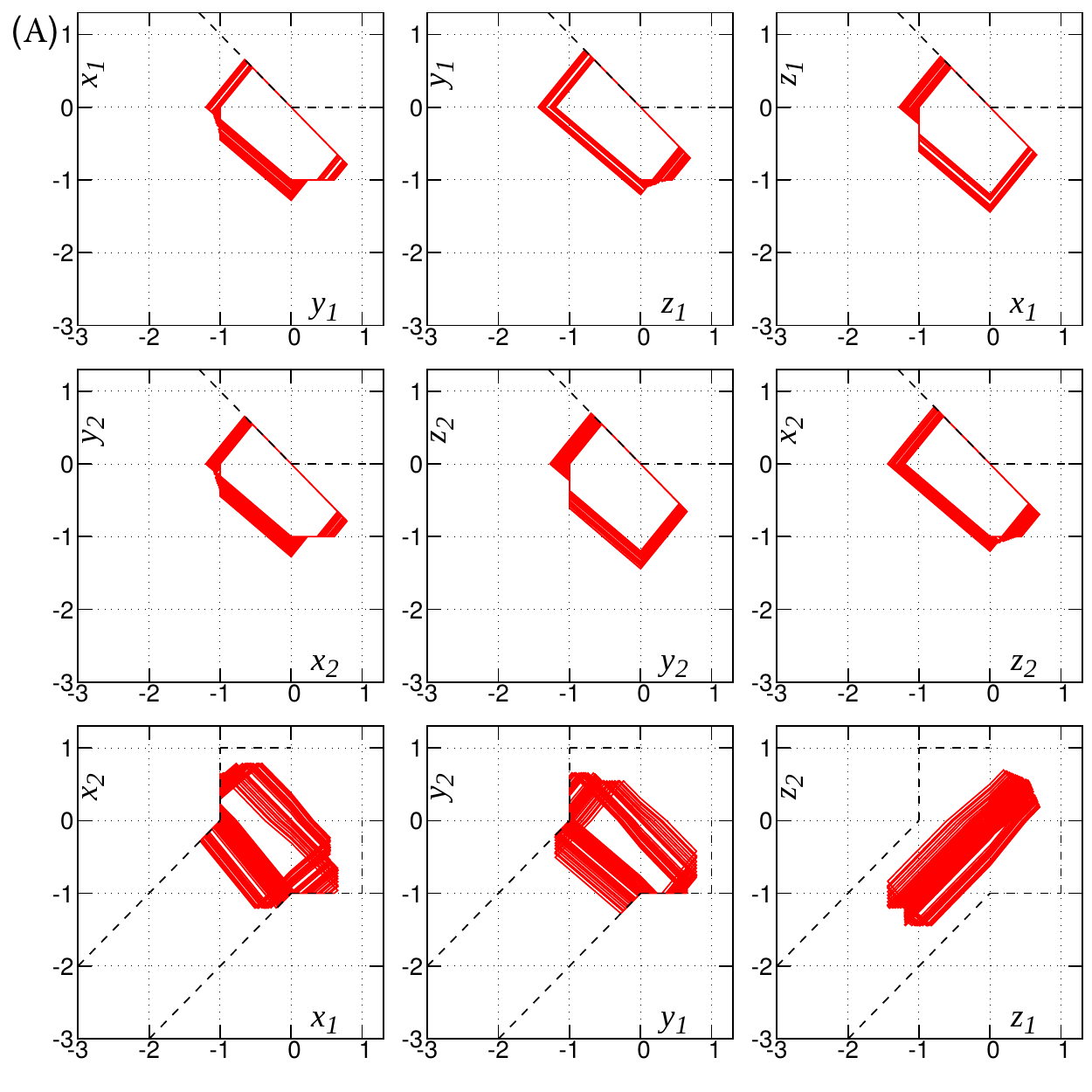}\hfill
\includegraphics[width=0.49\textwidth]{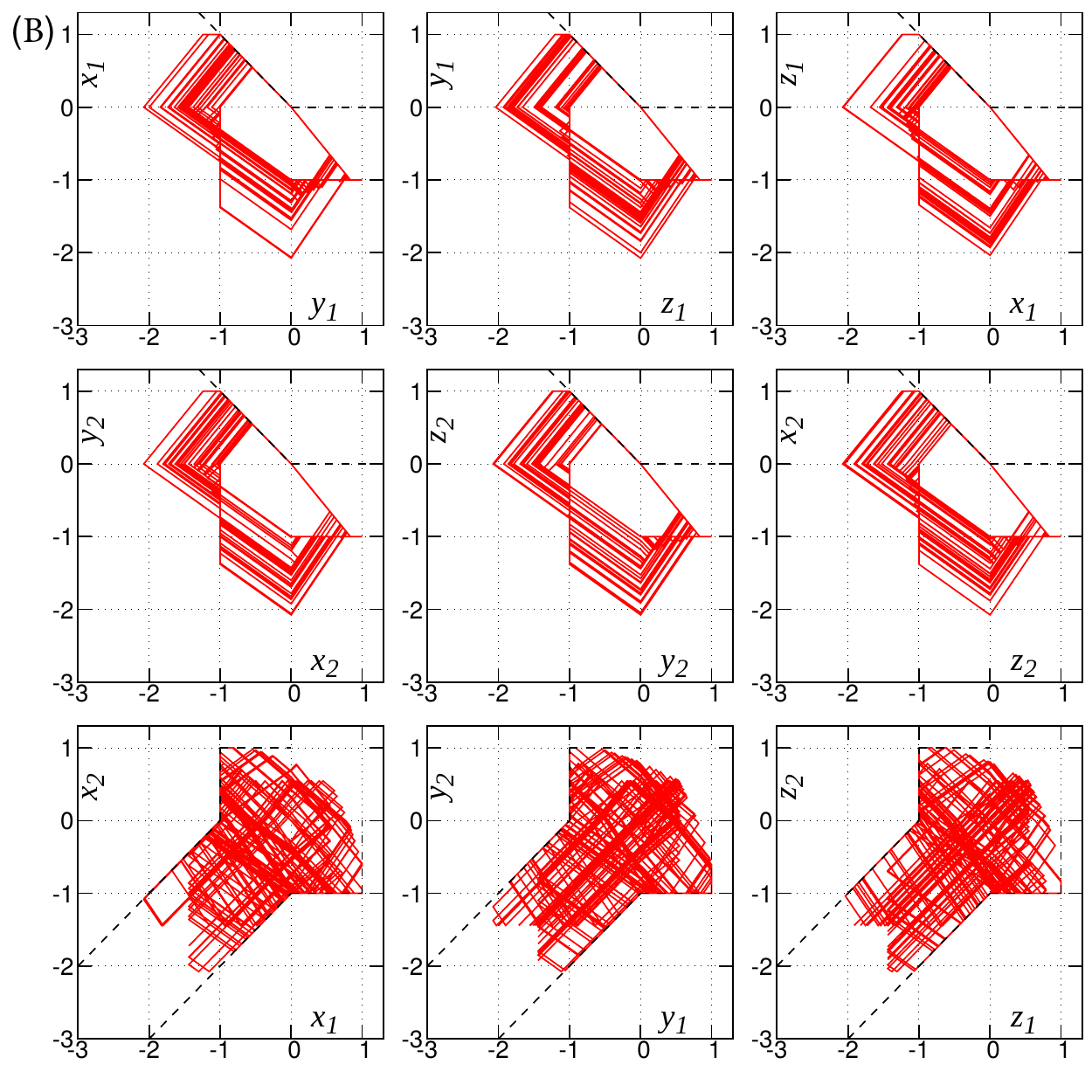}
\caption{Projections of a trajectory on different subplanes, for $\beta=1/6,\alpha=2.0$ (chaos of type Ax, panel (A)) and $\beta=1/6,\alpha=2.2$ (chaos of type B, panel (B)). Dashed lines show available domains according to Figs.~\ref{fig:funF},\ref{fig:domc}.}
\label{fig:trAB}
\end{figure}

Two types of chaotic regimes are illustrated 
in  Fig.~\ref{fig:trAB}. Chaos of type A (Fig.~\ref{fig:trAB}(A)) 
inherits symmetry of the periodic cycle of type A: there are 
three different attractors Aa, Ab, Ac according to renaming symmetry \eqref{eq:renam}. Chaotic regime of type B  (Fig.~\ref{fig:trAB}(B)) is fully symmetric,
both to renamings and to exchange.

Below we study chaos of type A at $\alpha=2$ in more details.

\begin{figure}[!htb]
\centering
\includegraphics[width=0.6\textwidth]{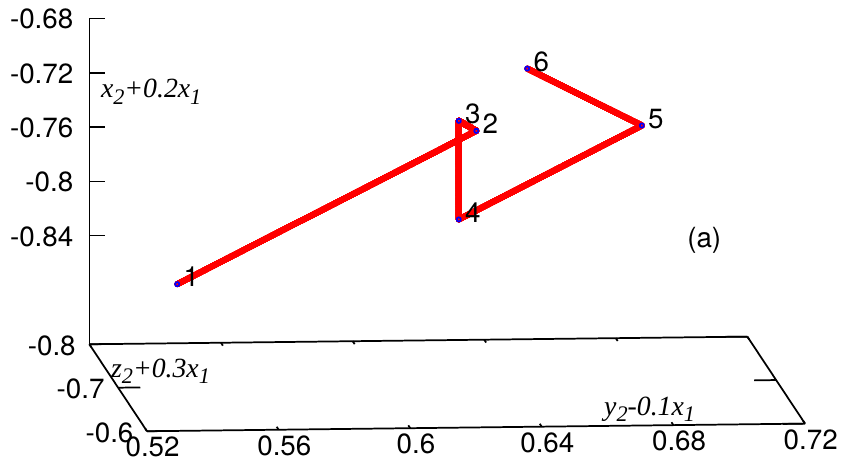}\hfill
\includegraphics[width=0.35\textwidth]{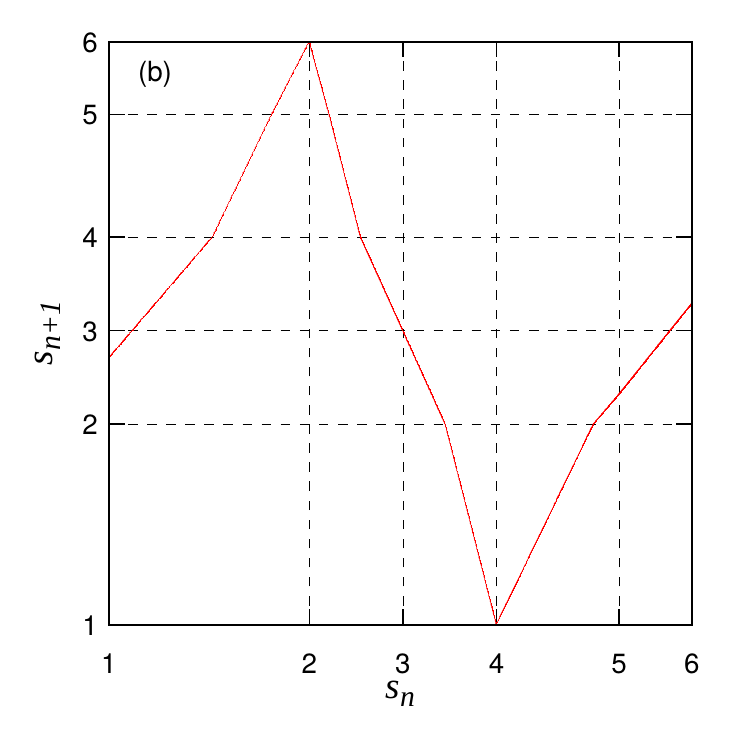}
\caption{Panel (a): Poincar\'e map and the borders of the segments.
Panel (b): Poincar\'e map in the one-dimensional parametrization
of the curve. Dashed lines are borders of the linear segments (marked at the axes with the same numbers as in panel (a)).}
\label{fig:pmI2}
\end{figure}

The starting point is a construction of a Poincar\'e map. As a hyperplane 
of section, we have chosen $y_1+0.5=0$, $\dot y_1<0$. At this section,
the value of the variable $z_1$ is also constant, because it is slaved by $y_1$.
The values of the other four variables $x_1,x_2,y_2,z_2$ are in general different, but there are segments where one of the variables is constant. Therefore,
to resolve all distinct pieces of the attractor, we show the 
 obtained sequence of dots is shown in Fig.~\ref{fig:pmI2}(a) in ``skew'' coordinates 
$x_2+0.2x_1,y_2-0.1x_1,z_2+0.3 x_1$. One can see that the attractor
consists of five straight intervals (the borders of these linear segments
are shown with markers and are marked with numbers). This suggests that the
attractor in the Poincar\'e section is one-dimensional and can be described with a one-dimensional map.

To construct this map, we parametrize the curve according to the Euclidean distance in the $6$-dimensional space.
The total length of five segments is $L=0.4523$. We denote the coordinate along the curve as $0\leq s\leq L$.
In Fig.~\ref{fig:pmI2}(b) we plot the one-dimensional Poincar\'e map $s_n\to s_{n+1}$. The points on the
axes and the grid are the borders of the linear segments. Remarkably, this map has no flat intervals, i.e. except
for a finite set of points  $|\frac{d s_{n+1}}{s_n}|>1$. This confirms chaoticity of the dynamics.

\subsection{Chaos of type B}

Here our approach is the same as in the characterization of chaos of type A above,
the parameter now is set to $\alpha=2.2$.

\begin{figure}[!htb]
\centering
\includegraphics[width=0.6\textwidth]{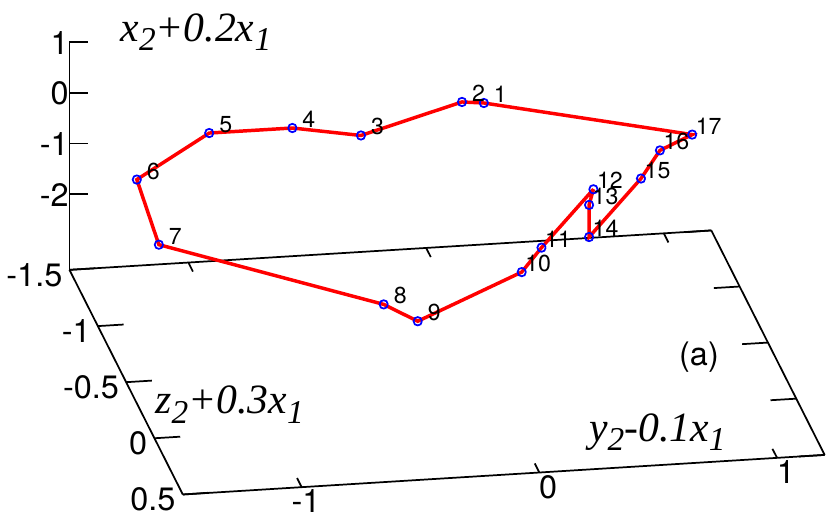}\hfill
\includegraphics[width=0.35\textwidth]{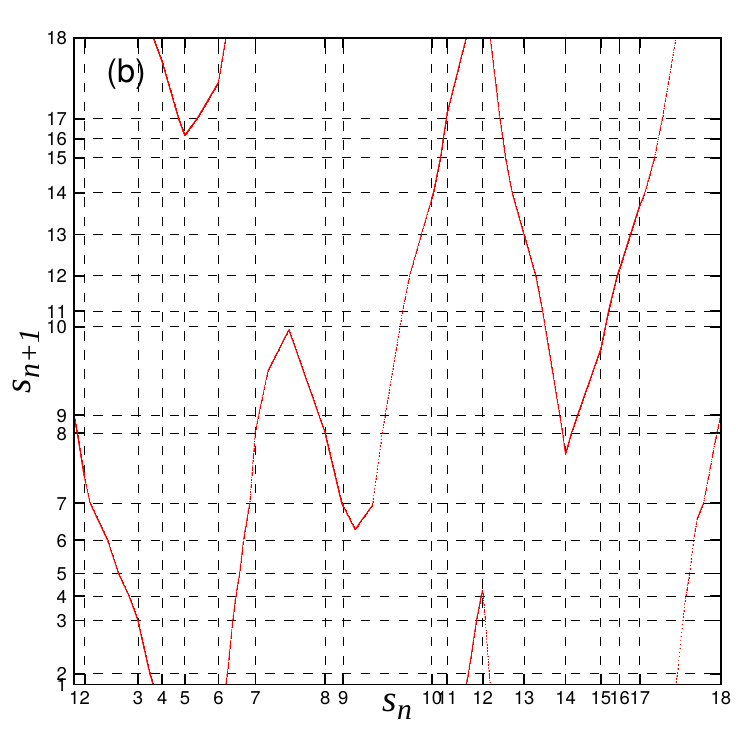}
\caption{Panel (a): Poincar\'e map in transformed coordinates. 
The borders of linear segments are denoted with numbers from $1$ to $17$.
Panel (b): Poincar\'e map in the one-dimensional parametrization
of the curve. Dashed lines are borders of the linear segments.}
\label{fig:lpm1}
\end{figure}

We use the same condition  $y_1=-0.5$, $\dot y_1<0$
for the Poincar\'e section, and plot
the obtained points in a three-dimensional space with coordinates $(y_2-0.1 x_1, z_2+0.3 x_1, x_2+0.2 x_1)$ (the value
of $z_1$ at this section is constant). Now the points form
not a one-dimensional piecewise-linear curve, but a closed curve,
consisting of 17 linear segments, see Fig.~\ref{fig:lpm1}(a).
This means that the attractor in the 5-dimensional
Poincar\'e map is one-dimensional, and 
topologically a circle.

Next, we parametrize the attractor  according to the Euclidean distance in the $6$-dimensional space.
The total length is $L=9.885$. The coordinate along the curve is $ s\in [0, L)$. Starting a trajectory on the attractor, and looking for a first return, we obtain a one-dimensional map  $s_n\to s_{n+1}$ plotted 
in Fig.~\ref{fig:lpm1}(b) we plot the one-dimensional Poincar\'e map $s_n\to s_{n+1}$. The points on the
axes and the grid are the borders of the linear segments (the same numbers as 
in Fig.~\ref{fig:lpm1}, point ``18'' is the same as ``1'').
This map is non-invertable and has 6 intervals of 
monotonicity. On all these intervals $|d s_{n+1}/d s_n|>1$,
what means that chaos is robust. However, there are regions where this
derivative is close to one (e.g., in segment 9), and presumably at the border of existence domain of chaos of type B, a stable cycle appears when at some segments the derivative
is less than one. We cannot, however, construct the corresponding
one-dimensional map for such a range of parameters, 
because we cannot obtain enough iteration points to form a closed curve like in Fig.~\ref{fig:lpm1}(a).

\section{Exploration of the continuous system close to the PC limit}
\label{sec:ecl}

In the previous sections, we explored the original system of coupled heteroclinic cycles \eqref{eq:2c} and its PC model \eqref{eq:cPC}. It is difficult to compare these findings because the PC model is valid in the limit $Q\to\infty$ and
the explored in Section~\ref{sec:nexp} values $Q\lesssim 40$ are far from this limit. Therefore, in this section we present simulations at rather large $Q=200$
(what corresponds to the coupling $D=1.38\cdot 10^{-87}$, at which numerical simulations are still reliable). Below in this section we fix $\beta=1/6$ and vary $\alpha$, to be able to compare the observed attractors with those of Section~\ref{sec:plach}.

\begin{figure}[!htb]
\centering
\includegraphics[width=0.48\textwidth]{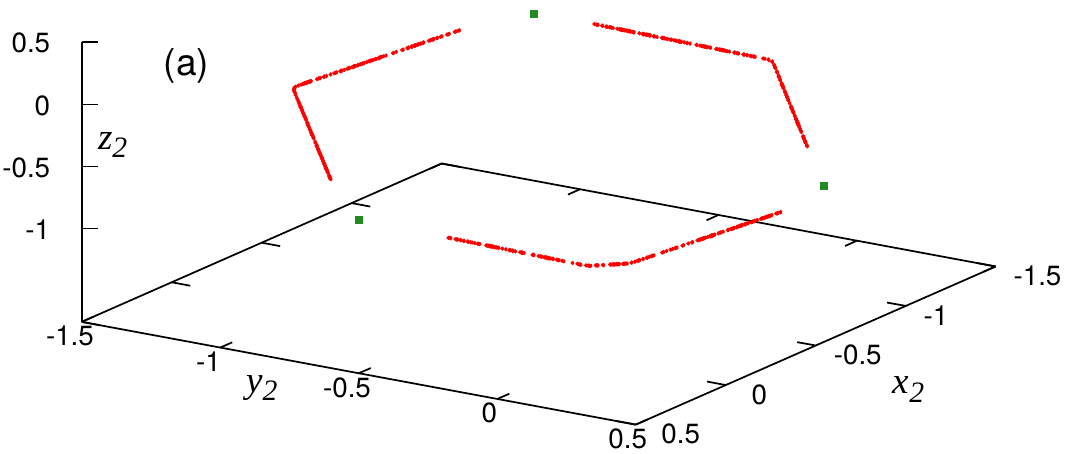}\hfill
\includegraphics[width=0.48\textwidth]{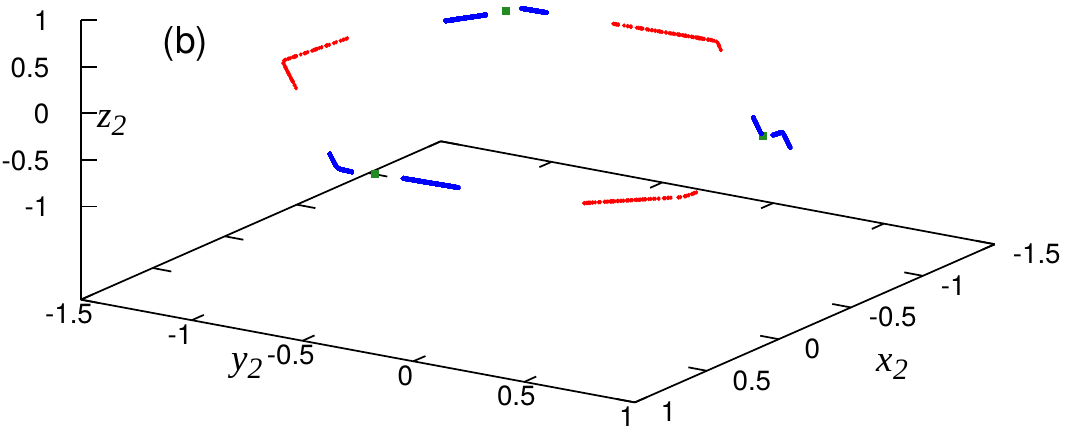}\\
\includegraphics[width=0.48\textwidth]{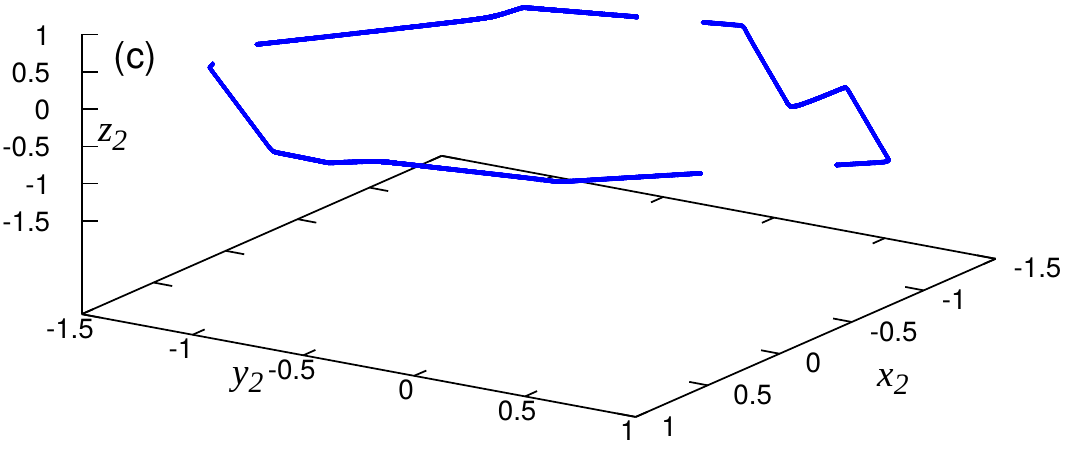}\hfill
\includegraphics[width=0.48\textwidth]{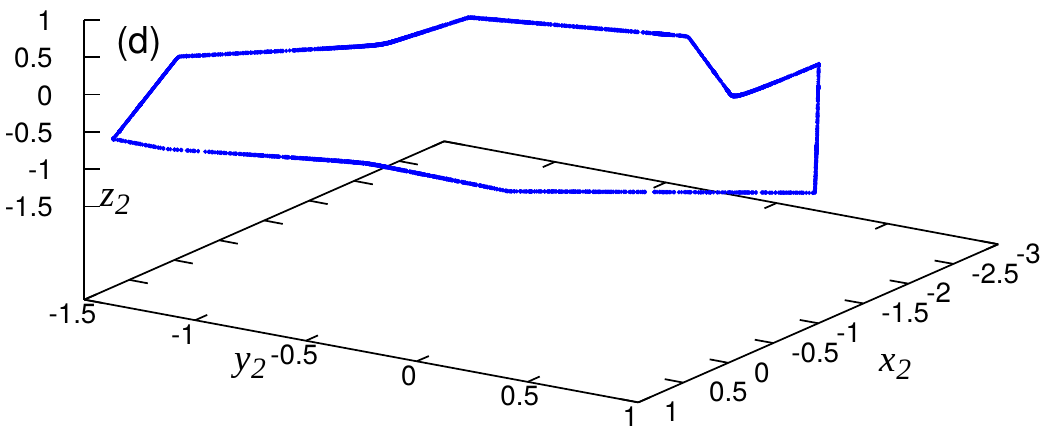}
\caption{Poincar\'e maps at section $y_1=-0.5;\;\dot y_1<0$. Points at this
section are shown in coordinates $x_2,y_2,z_2$. (a) $\alpha=1.86$. Here, only stable symmetric orbits 
 (green squares) and neutral periodic orbits (red dots) are present. (b) $\alpha=1.93$. Here, stable symmetric orbits 
 (green squares), neutral periodic orbits (red dots), and
 small-scale chaos (blue dots) are present.  
 (c) $\alpha=2.08$. Here,  the asymmetric large-scale attractor (blue dots) is present. (d) $\alpha=2.70$. Here, the symmetric large-scale attractor is present.
}
\label{fig:rand1}
\end{figure}

Numerical exploration at each value of $\alpha$ was performed as follows:
for a large set of randomly chosen initial conditions,
the final state (``attractor'') after a long transient $10^6$ was analyzed. The following attractors are observed (we depict them in the Poincar\'e plots in Fig.~\ref{fig:rand1}):

\textbf{Stable synchronous periodic orbits on manifolds M1,M2,M3.} These attractors are observed 
in the range $1.85\leq \alpha \leq 2.07$. In Figure \ref{fig:rand1}(a,b)
these periodic orbits
are depicted with green squares.

\begin{figure}[!htb]
\centering
\includegraphics[width=0.6\textwidth]{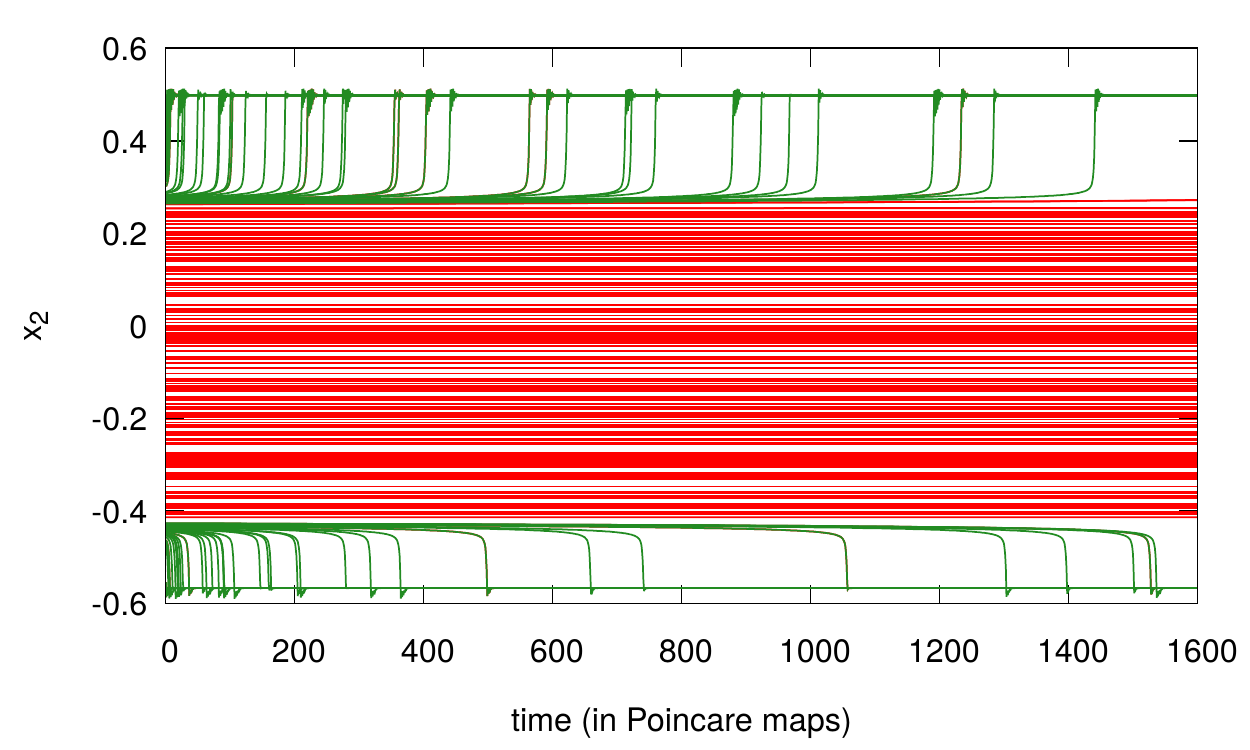}
\caption{Evolution of neutral periodic orbits (red lines) and of trajectories close to them (green lines). The horizontal axis shows discrete times for the Poincar\'e map, in real time this corresponds to $\approx 1.2\cdot 10^6$. One can see practically no evolution of the red trajectories  in the bulk, and convergence of
green trajectories starting at the boundaries 
to the stable synchronous orbits. This picture is very similar to one reported for the oscillators coupled with a dead zone coupling (Fig 3(b1) in \cite{ashwin2021dead}). 
}
\label{fig:npo}
\end{figure}

\textbf{``Neutral periodic orbits''}. These orbits, which are similar to the periodic 
orbits of type A in the PC model, appear to build three families; they are 
characterized by
a continuous parameter.   In Figure \ref{fig:rand1}(a,b) 
these orbits are depicted with red dots.   These attractors are observed 
in the range $1.85\leq \alpha \leq 1.98$. Remarkably, for larger values of parameter $\alpha$,
the range of these 
orbits becomes smaller.
Numerical limitations do not allow concluding, whether these
orbits are true neutral ones, like the theory of dead zones
predicts~\cite{ashwin2019state,ashwin2021dead}, or they
just evolve so slowly that within the explored time interval they look like stationary. To illustrate this, we plot in Fig.~\ref{fig:npo} the evolution of these orbits together with neighboring
ones that converge eventually to real attractors - stable symmetric periodic orbits.

\begin{figure}[!htb]
\centering
\includegraphics[width=0.49\textwidth]{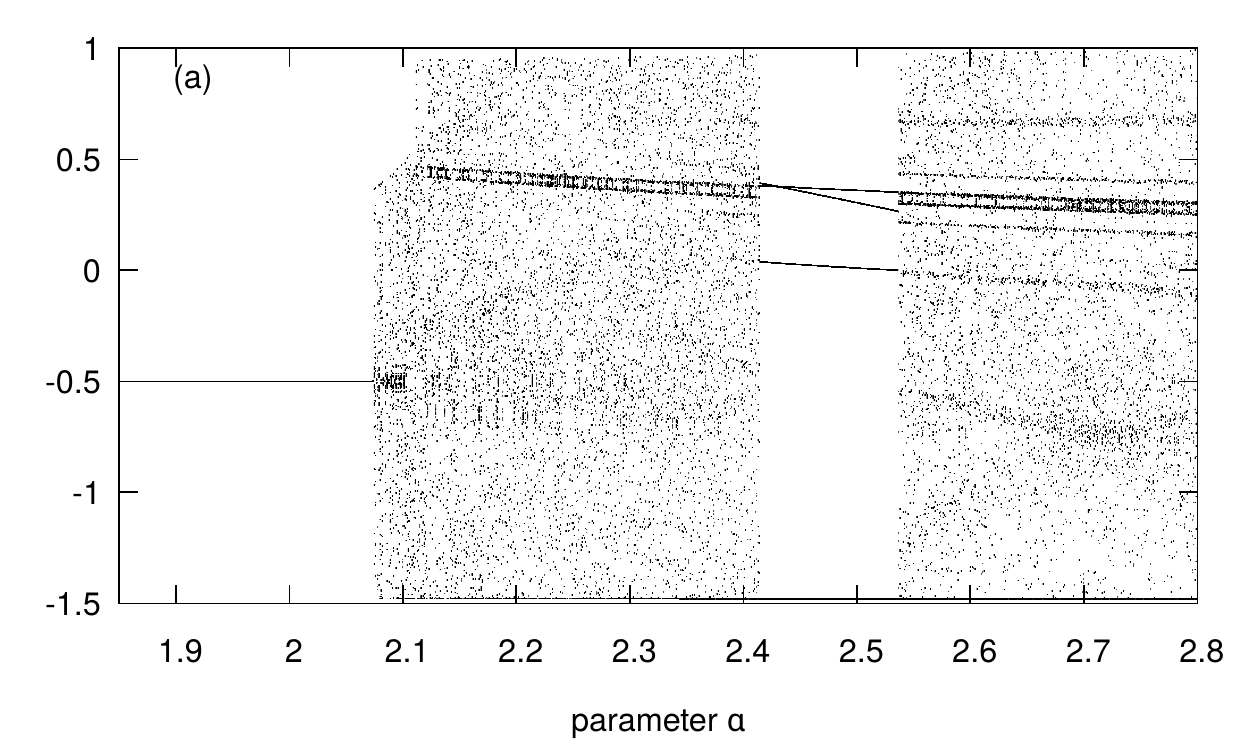}\hfill
\includegraphics[width=0.49\textwidth]{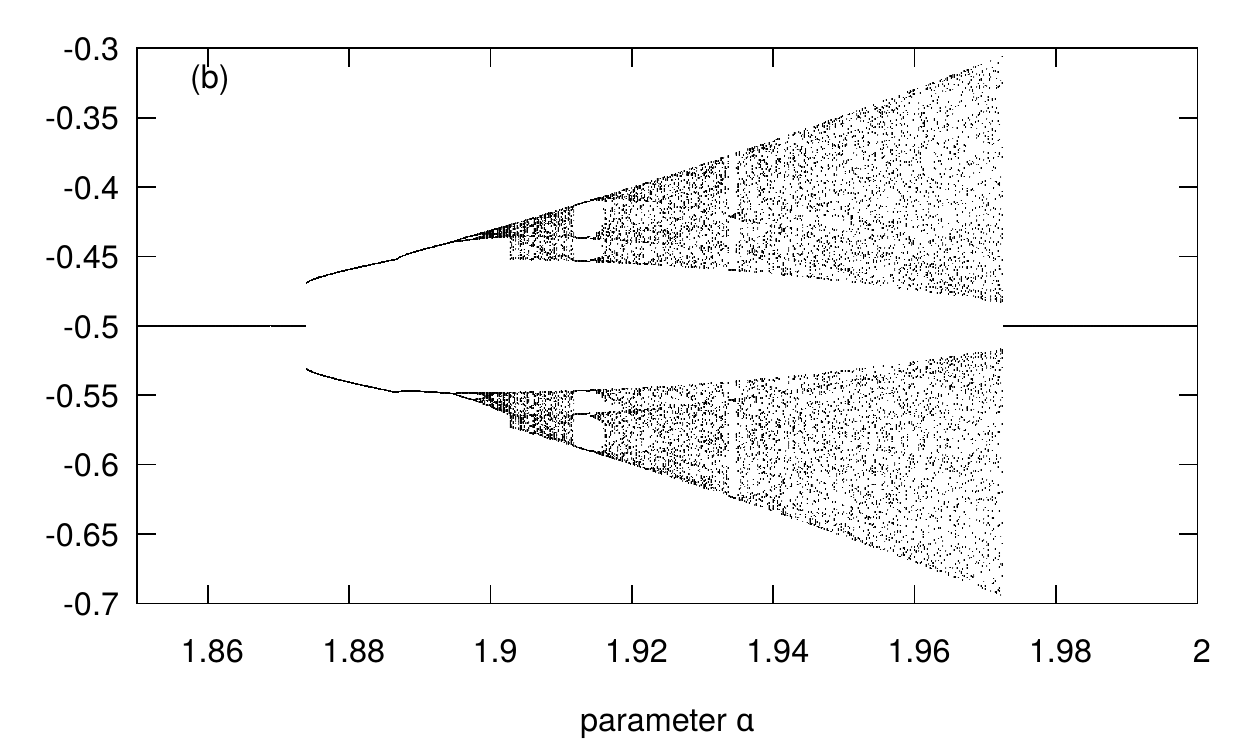}
\caption{Panel (a): Bifurcation diagram
(Poincar\'e maps at section $y_1=-0.5;\;\dot y_1<0$) for the large-scale chaotic attractor.
Panel (b): Bifurcation diagram for the small-scale attractor.
Because this branch is isolated, 
to obtain it we started within chaos at $\alpha=1.97$ and varied $\alpha$ with small steps in both directions until this attractor disappeared. On both panels the fixed point at $y_2=-0.5$
is a symmetric stable periodic orbit.
}
\label{fig:bdsc}
\end{figure}

\textbf{Large-scale asymmetric chaos and large-scale symmetric chaos} These attractors are similar to chaos of types A and B, respectively, in the PC model,  they are observed for $\alpha > 2.07$, and the transition from asymmetric to symmetric one occurs at $\alpha\approx 2.11$. 
In Figure \ref{fig:rand1}(c,d)  these orbits are depicted with blue dots. Figure \ref{fig:bdsc}(a) show the bifurcation diagram
of these attractors, obtained by slowly decreasing parameter $\alpha$ from large values. Inside these chaotic states, there are periodic windows.

\textbf{Small-scale chaos} 
These three symmetric branches of solutions appears to be isolated from other
attractors, they are  observed for $1.87 \leq \alpha\leq 1.97$ and are shown in Figures \ref{fig:rand1}(b) with blue dots.  
with blue dots. Additionally, a bifurcation diagram for one of such attractors is shown in Fig.~\ref{fig:bdsc}(b).

To conclude this section, we compare the regimes in the continuous system at $Q=200$ with those in the PC model:
\begin{itemize}
\item Large scale chaos (asymmetric and symmetric) for $Q=200$ corresponds 
to chaotic  regimes of types A and B in the PC model. However, the ranges of  existence in dependence on parameter $\alpha$ are different. 
\item The property of having a continuous set of periodic orbits in the PC model corresponds to the observation of 
such periodic orbits also for $Q=200$, although here possibly the orbits are slightly non-neutral with an
extremely slow evolution toward stable symmetric orbits.
However, this slow evolution
can be hardly followed for the available length scale and for a limited precision of calculations.
\item In the PC model the symmetric solutions on manifolds M1, M2, M3 appear unstable,
while in the continuous model for $Q=200$ they are stable
in a certain range of the parameter $\alpha$.
\item Small-scale chaos in the continuous model 
at $Q=200$ does not have a counterpart in the PC model.
\item In the PC model for large $\alpha$, a stable periodic orbit is observed, while such an orbit 
appears unstable (if it exists) at $Q=200$, where chaos is observed at large values of parameter $\alpha$.
\end{itemize}

\section{Conclusion}
\label{sec:concl}

In this paper, we explored  coupled heteroclinic cycles, rotating in opposite directions. This makes their dynamics completely different to that of coupled heteroclinic cycles rotating in the same direction \cite{PhysRevE.85.016215},
or to coupling of a heteroclinic and a usual limit cycle~\cite{TACHIKAWA2007374}. In the present setup, the two systems necessarily come to states where there is  an interaction between one large and one small variable. This makes the effect of even a very small coupling enormous. In our numerical simulations of the full system, we observed that chaos persist for very small coupling strengths.  An increase 
of coupling strength results in a series of transitions related to change of symmetry of the chaotic attractors; finally, an inverse period-doubling transition to a stable periodic cycle occurs. 

Existence of chaos in the limit of small coupling allows for a construction of a proper analytical model valid in this limit. The model constructed is based 
on a piecewise-constant representation of r.h.s. of the equations in transformed (logarithmic) variables. In many previous studies, such an approximation of dynamical equations led to a possibility of an analytic representation of the dynamics via explicit construction of a Poincar\'e map \cite{Pikovsky-Rabinovich-78,Kijashko-Pikovsky-Rabinovich-80,Pikovsky-Rabinovich-81,guckenheimer2006chaotic}. In other cases, already the basic
model is formulated as a piecewise-linear one, what enabled for an explicit construction of chaotic solutions there~\cite{belykh2019lorenz,belykh2020bifurcations,belykh2021sliding}. Mostly
close to our consideration is the analysis of chaos in complex 
heteroclinic connections for a Bianchi IX model in cosmology~\cite{bogoyavlenskii1973singularities,bogoyavlenskii1976homogeneous},
where an asymptotic one-dimensional map has been constructed for a complex heteroclinic cycle. 
Remarkably, the piecewise-constant model is scale-invariant: the dynamics
does not depend on the coupling parameter, what suggests that it really captures the singular limit of zero coupling. 

Unfortunately, the dynamics in the full 6-dimensional phase space in this model appears too complex to represent it completely analytically (like in other systems with the piecewise-constant dynamics, demonstrating chaos~\cite{schurmann1995entropy,peters2003hybrid,blank2004switched}). Therefore, we followed a numerical approach, where conditions and borders of available domains were determined numerically, while the pieces of trajectories between these turning points were drawn from the exact equations. In this way, we constructed Poincar\'e maps, which appeared to be expanding non-invertable
one-dimensional maps. 

A peculiar property of the interaction in the PC limit is that there are domains
of variables where the coupling vanishes exactly, so that the coupling works
only at the boundaries of these domains. Oscillatory systems with such a property
have been recently introduced by using a concept of dead zones (zones without coupling) \cite{ashwin2019state,ashwin2021dead}. The coupled heteroclinic cycles 
in the PC limit appear to be an example of a dead zone system. This explains a rather unusual property of existence of a family of locked cycles with phase shifts from some interval. We further demonstrated that this property is also observed in the continuous system, although we hypothesize that this observation may be limited to finite although very large time intervals, because in reality the coupling does not vanish but is exponentially small (cf.~\cite{Bolotov-Osipov-Pikovsky-16}). 

Above, we restricted ourselves to a symmetric situation, where
the parameters inside the cycles are the same and the coupling is also symmetric. We expect that the constructed chaotic attractor is robust under breaking of these conditions, but this hypothesis has to be checked in future studies. 

In this paper, we considered two coupled heteroclinic cycles, it appears natural
to extend this study to heteroclinic networks. We stress here that ``heteroclinic networks'' can be understood in different senses. One can consider a lattice or a network, where on each cite there is a heteroclinic cycle, and these cycles are coupled via links with other cycles like in Eqs.~\eqref{eq:2c}, see \cite{10.1143/PTP.109.133}. Our approach could be
generalized for such a setup, if one assumes different rotation directions at different nodes. In another interpretation,
a network consists of many nodes $\xi_1,\xi_2,\xi_3,\ldots,\xi_n$ with saddle connections $\xi_k\to\xi_m$ between them \cite{ashwin2020almost}. In the context of our study above, one could look at a weak coupling of two such networks, where at least some connections are reverted.

\section*{Acknowledgements}
The authors acknowledge the partial financial support by the Israel Science Foundation (grant No. 843/18).
AP acknowledges support by the Laboratory
of Dynamical Systems and Applications NRU HSE of the Russian
Ministry of Science and Higher Education (Grant No. 075-15-2019-
1931).
We thank Michael Zaks for fruitful discussions.
\bibliography{paper_chc.bib}
\appendix
\section{Hopf bifurcation of a coexistence point on a synchronous manifold}
\label{sec:hbif}
Below we consider the bifurcation of the limit cycle on the invariant manifold M1 defined by relations \eqref{eq:mf1}, which appears due to the oscillatory instability of the ``coexistence point" $(a,a,a)$, $a=(\alpha+\beta+1)^{-1}$ (see Section 4.2).

It is convenient to transform system \eqref{eq:m1} using the following rescaling of variables:
$$t=a^{-1}\tau,\;D=ad,\;u=a(1+U),\;v=a(1+V),\;w=a(1+W).$$
We obtain:
$$\frac{dU}{d\tau}+U+\alpha V+\beta W=-U(U+\alpha V+\beta W),$$
\begin{equation}
    \frac{dV}{d\tau}+\beta U+(1+d)V+(\alpha-d)W=-V(V+\alpha W+\beta U)
    \label{eqapp1}
\end{equation}
    $$\frac{dW}{d\tau}+\alpha U+(\beta-d)V+(1+d)W=-W(W+\alpha U+\beta V).$$

In order to derive the amplitude equation governing the evolution of small-amplitude disturbances near the coexistence point in the vicinity of the instability threshold point, 
\begin{equation}
d=d_0+\epsilon^2d_2,\;\epsilon\ll 1,
\label{eqapp2}
\end{equation}
we apply the multi-scale approach. We present the variables $U$, $V$ and $W$ as functions of two variables, $\tau_0=\tau$ and $\tau_2=\epsilon^2\tau$,
$$U=U(\tau_0,\tau_2),\;V=V(\tau_0,\tau_2),\;W=W(\tau_0,\tau_2),$$
and replace the differentiation operator with
\begin{equation}
    \frac{d}{d\tau}=\frac{d}{d\tau_0}+\epsilon^2\frac{d}{d\tau_2}.
    \label{eqapp3}
\end{equation}
    Here $\tau_0$ corresponds to fast oscillations of the variables, and $\tau_2$ describes the ``slow" evolution of the amplitude and the phase of oscillations near the instability threshold.
    The variables are presented as the series,
\begin{equation}
    (U,V,W)=\epsilon(U^{(1)},V^{(1)},W^{(1)})+\epsilon^2(U^{(2)},V^{(2})+\epsilon^3(U^{(3)},V^{(3)},W^{(3)})+\ldots
    \label{eqapp4}
\end{equation}

At the leading order we obtain the linear system
$$\frac{dU^{(1)}}{d\tau_0}+U^{(1)}+\alpha V^{(1)}+\beta W^{(1)}=0,$$
\begin{equation}
    \frac{dV^{(1)}}{d\tau_0}+\beta U^{(1)}+(1+d_0)V^{(1)}+(\alpha-d_0)W^{(1)}=0,
    \label{eqapp5}
\end{equation}
$$\frac{dW^{(1)}}{d\tau_0}+\alpha U^{(1)}+(\beta-d_0)V^{(1)}+(1+d_0)W^{(1)}=0.$$
The expression for the growth rates,
$$\sigma_{\pm}=\frac{\alpha+\beta-2}{2}-d_0\pm i\sqrt{\frac{3(\alpha-\beta)^2}{4}-d_0^2},$$
which is equivalent to \eqref{eq:scp} up to rescaling of variables, determines the instability threshold
$$d_0=\frac{\alpha+\beta-2}{2}$$
and the frequency of oscillations,
$$\omega_0=\mbox{Im}\sigma=\frac{1}{2}\sqrt{3(\alpha-\beta)^2-(\alpha+\beta-2)^2}.$$

The solution of \eqref{eqapp5} is
$$(U^{(1)},V^{(1)},W^{1})=A(\tau_2)e^{i\omega_0\tau_0}(U_1^{(1)},V_1^{(1)},W_1^{(1)})+A(\tau_2)^*e^{-i\omega_0\tau_0}(U_1^{(1)*},V_1^{(1)*},W_1^{(1)*}),$$
where $A(\tau_2)$ is a complex amplitude function, which is still unknown. Choosing $U_1^{(1)}\equiv 1$, we obtain:
$$U_1^{(1)}=1,\;V_1^{(1)}=-\frac{1-\beta+i\omega_0}{\alpha-\beta},\;
W_1^{(1)}=-\frac{\alpha-1-i\omega_0}{\alpha-\beta}.$$
 
At the second order, we find
$$(U^{(2)},V^{(2)},W^{(2)}=|A(\tau_2)|^2(U_0^{(2)},V_0^{(2)},W_0^{(2)})+$$
$$A^2(\tau_2)e^{2i\omega_0\tau_0}(U_2^{(2)},V_2^{(2)},W_2^{(2)})+A^{*2}(\tau_2)e^{-2i\omega_0\tau_0}(U_2^{(2)*},V_2^{(2)*},W_2^{(2)*}),$$
where
$$U_0^{(2)}=\frac{\alpha+\beta-2}{\alpha+\beta+1}.\;V_0^{(2)}=-\frac{(\alpha+\beta-2)(2\beta-1)}{(\alpha+\beta+1)(\alpha-\beta)},\;W_0^{(2)}=\frac{(\alpha+\beta-2)(2\alpha+1)}{(\alpha+\beta+1)(\alpha-\beta)};$$
$$U_2^{(2)}=\frac{1}{12\omega_0^2}[\alpha^2-10\alpha\beta+\beta^2+8\alpha+8\beta-8-2i\omega_0(\alpha+\beta-2)]+\frac{1}{6\omega_0^2}\left(\frac{\alpha+\beta-2}{\alpha-\beta)}\right)^2[2\alpha\beta-\alpha-\beta+2i\omega_0(\alpha+\beta)],$$
$$V_2^{(2)}=-\frac{1}{12\omega_0^2}[2\alpha^2-5\alpha\beta-\beta^2+\alpha+7\beta-4+2i\omega_0(-2\alpha+\beta+1)]+$$
$$\frac{1}{6\omega_0^2}\left(\frac{\alpha+\beta-2}{\alpha-\beta}\right)^2\frac{\alpha+\beta-2-i\omega_0)(2\alpha^2-7\alpha\beta+\beta^2+4\alpha+3\beta-3-2i\omega_0\beta)}{\alpha+\beta+1+2i\omega_0},$$
$$W_2^{(2)}=\frac{1}{12\omega_0^2}[\alpha^2+5\alpha\beta-2\beta^2-\beta-7\alpha+4-2i\omega_0(\alpha-2\beta+1)]+$$
$$\frac{1}{6\omega_0^2}\left(\frac{\alpha+\beta-2}{\alpha-\beta}\right)^2\frac{(\alpha+\beta-2-i\omega_0)(\alpha^2-7\alpha\beta+2\beta^2+3\alpha+4\beta+3-2i\omega_0\alpha)}{\alpha+\beta+1+2i\omega_0}.$$

At the third order in $\epsilon$, following the idea of the method of multiple scales, we demand the solvability of the obtained system in the class of functions periodic in $\tau_0$ (i.e., we eliminate the secular terms). The solvability condition gives the amplitude equation 
$$\frac{dA}{d\tau_2}=sd_2A-\kappa|A|^2A$$
that describes the slow change of the amplitude function. Here 
$$s=\left(\frac{\partial\sigma_+}{\partial d}\right)_{d=d_0}=-1-i\frac{\alpha+\beta-2}{2\omega_0}.$$
The expression for $\kappa(\alpha,\beta)$ is rather cumbersome, and we do not write it here. Let us present that expression in the limit of small $\delta\equiv d_0=(\alpha+\beta-2)/2$:
$$\kappa=2i\sqrt{3}(\alpha-1)+2\delta(4-i\sqrt{3}\alpha)+O(\delta^2).$$

One can see that for arbitrary small $\delta>0$, Re$\kappa>0$. Thus, there is a direct Hopf bifurcation: the small stable solution with $|A|^2=d_2\mbox{Re}s/\mbox{Re}\kappa$ exists in the region $d_2<0$, where the coexistence point is unstable.

Note that in the case $\alpha+\beta=2$, $d_0=0$ (i.e., for system \eqref{eq:hc1}, Re$s=0$, hence we do not observe the standard Hopf bifurcation. Indeed, it is known that for $\alpha+\beta=2$, system \eqref{eq:hc1} has an infinite number of periodic solutions corresponding to oscillations with different amplitudes.

\end{document}